\documentclass[review]{elsarticle}
\usepackage{lineno,hyperref}
\modulolinenumbers[5]
\usepackage[utf8]{inputenc}
\usepackage{amsmath, amssymb}
\usepackage{mathtools}
\usepackage{placeins}
\usepackage{longtable}
\usepackage{graphicx}
\usepackage{bbm}
\PassOptionsToPackage{numbers, compress}{natbib}
\usepackage{natbib}
\usepackage{lineno}
\usepackage{setspace}
\usepackage{xcolor}
\usepackage{makecell}
\renewcommand{\d}{\mathrm{d}}
\newcounter{oldtocdepth}
\newcommand{\hidefromtoc}{%
  \setcounter{oldtocdepth}{\value{tocdepth}}%
  \addtocontents{toc}{\protect\setcounter{tocdepth}{-10}}%
}
\newcommand{\unhidefromtoc}{%
  \addtocontents{toc}{\protect\setcounter{tocdepth}{\value{oldtocdepth}}}%
}

\newcommand{\F}{\mathcal{F}}

\journal{Journal}

\begin{document}

\begin{frontmatter}

\title{Automated Dissipation Control for Turbulence Simulation with Shell Models}

\author[a]{Ann-Kathrin Dombrowski}
\ead{a.dombrowski@tu-berlin.de}

\author[a,b,c,d]{Klaus-Robert M\"uller\corref{corr}}
\ead{klaus-robert.mueller@tu-berlin.de}

\author[e]{Wolf Christian M\"uller\corref{corr}}
\ead{wolf-christian.mueller@tu-berlin.de}

\cortext[corr]{Corresponding author.}
\address[a]{Machine Learning Group, TU-Berlin, Germany}
\address[b]{Department of Artificial Intelligence, Korea University, Seoul, Korea}
\address[c]{Max Planck Institute for Informatics, Saarbrücken, Germany}
\address[d]{BIFOLD -  Berlin Institute for the Foundations of Learning and Data,  TU-Berlin, Germany}
\address[e]{Plasma Astrophysics, TU-Berlin, Germany}

\begin{abstract}

The application of machine learning (ML) techniques, especially neural networks, has seen tremendous success at processing images and language. This is because we often lack formal models to understand visual and audio input, so here neural networks can unfold their abilities as they can model solely from data.
In the field of physics we typically have models that describe natural processes reasonably well on a formal level. Nonetheless, in recent years, ML has also proven useful in these realms, be it by speeding up numerical simulations or by improving accuracy.
One important and so far unsolved problem in classical physics is understanding turbulent fluid motion.
In this work we construct a strongly simplified representation of turbulence by using the Gledzer-Ohkitani-Yamada (GOY) shell model. With this system we intend to investigate 
the potential of ML-supported and physics-constrained small-scale turbulence modelling. Instead of standard supervised learning we propose an approach that aims to reconstruct statistical properties of turbulence such as  the self-similar inertial-range scaling, where
we could achieve encouraging experimental results. Furthermore we discuss pitfalls when combining machine learning with differential equations.

\end{abstract}

\end{frontmatter}
%\linenumbers

\hidefromtoc
\section{Introduction}
Turbulence is irregular, unsteady and seemingly chaotic fluid motion. 
Opportunities to witness turbulence are plentiful: the water in a fast streaming river or waterfall, the smoke of a fire, the air behind a flying plane or the plume formed by a rocket motor are all examples of turbulent flows.
For many technical applications it is important to understand, predict and control turbulent motion, so history of research and progress on analyzing  turbulence is vast and rich ~\cite{wilcox1998turbulence,frisch1995turbulence,tennekes2018first,schlichting2016boundary,batchelor1953theory,turner1979buoyancy,monin2013statistical,pope2001turbulent}. Nonetheless a complete understanding of its spatio-temporal nonlinear behaviour remains a great challenge in physics (see e.g. ~\cite{pope2001turbulent} for a detailed description of the fundamentals of turbulent flows).
In the following, we will refer to three-dimensional hydrodynamic Navier-Stokes turbulence.
Traditionally, strongly subsonic gas- and hydrodynamic turbulence is studied using the incompressible Navier-Stokes equations 
\begin{align}
\frac{\partial \boldsymbol{u}}{\partial t} + (\boldsymbol{u} \cdot\nabla)\boldsymbol{u} &= -\nabla p+\mathsf{Re}^{-1}\Delta\boldsymbol{u} + \boldsymbol{f}\\
\nabla\cdot\boldsymbol{u} &= 0 ,
\label{eq:Navier_Stokes}
\end{align}
given here in dimensionless form with the velocity $\boldsymbol{u}$, the pressure $p$, and the Reynolds number  $\mathsf{Re}=L_0U_0/\nu$ that is defined by a characteristic length, $L_0$, velocity, $U_0$, and the kinematic viscosity $\nu$. The vector $\boldsymbol{f}$ denotes a force-field that drives the turbulent flow on large spatial scales.
The Reynolds number roughly determines the ratio of the largest spatial scale of turbulence, $L_0$, and the smallest one, $\eta$. The Kolmogorov length $\eta$ estimates the scale where the nonlinear spectral flux due to the energy cascade becomes comparable to the dissipative losses. As atmospheric and oceanic flows easily reach $\mathsf{Re} \gtrsim 10^8$, a fully resolved and spatially discretized representation of realistic turbulence requires an enormous spectral bandwidth,
$L_0/\eta \sim \mathsf{Re}^{3/4}$~\cite{annurevscalings}.  
 
In most cases a solution to the Navier-Stokes equations cannot be derived analytically and is thus computed using numerical methods.
For the reason given above, Direct Numerical Simulations (DNS)~\cite{pope2001turbulent}
at realistic Reynolds numbers are computationally not feasible without further approximation.
%since the grid on which the solution is computed needs to be very fine in order to capture %turbulent motion at the smallest scales, which can be roughly estimated as $\sim L_0 %\mathsf{Re}^{-3/4}$ (spatially) and $\sim t_0 \mathsf{Re}$ (temporally) with $t_0=L_0/U_0$.
In these cases Large-Eddy Simulations (LES)~\cite{sagaut2001LES, pope2001turbulent} can be useful.
The LES equations are obtained by applying a linear operator that acts as a low pass filter to the Navier-Stokes equations~(\ref{eq:Navier_Stokes}). Written in component notation and employing the Einstein summation convention we then get:
\begin{align}
\frac{\partial \bar{u}_i}{\partial t} + \frac{\partial(\overline{u_i u_j})}{\partial x_j} &= -\frac{\partial \bar{p}}{\partial x_i} + \nu \frac{\partial^2 \bar{u}_i}{\partial x_j \partial x_j} + \bar{f}_i, \\
\frac{\partial \bar{u}_j}{\partial x_j} &= 0 
\end{align}
where $\bar{\bullet}$ denotes a filtered quantity.
These differential equations describe the dynamics of a spatially filtered velocity field, i.e. the large-scale flow. Thus the numeric calculations can be performed on a coarser grid.
Due to the non-linear nature of the advection term $\frac{\partial(u_i u_j)}{\partial x_j}$, its filtered version $\frac{\partial(\overline{u_i u_j})}{\partial x_j}$ is still dependent on the original velocity field. This is the consequence of the fact that  nonlinear interactions do not exclusively occur among filtered fluctuations which are resolved on the coarser grid but also between filtered fluctuations and unresolved components of the velocity field that have been eliminated by the filter operation.
Those latter interactions give rise to a subgrid-scale stress $\tau_{ij}=\overline{u_iu_j}-\overline{u}_i\overline{u}_j$ that can only be determined with the   
unfiltered velocity field:

\begin{equation}
%\frac{\partial \bar{u_i}}{\partial t} + \frac{\partial(\overline{u_i u_j})}{\partial x_j} + \frac{\partial(\bar{u}_i \bar{u}_j)}{\partial x_j} &= -\frac{\partial \bar{p}}{\partial x_i} + \nu \frac{\partial^2 \bar{u}_i}{\partial x_j \partial x_j} + \bar{f_i} +  \frac{\partial(\bar{u}_i \bar{u}_j)}{\partial x_j}\\
% \frac{\partial \bar{u}_i}{\partial t} + \frac{\partial(\bar{u}_i \bar{u}_j)}{\partial x_j} &= -\frac{\partial \bar{p}}{\partial x_i} + \nu \frac{\partial^2 \bar{u}_i}{\partial x_j \partial x_j} + \bar{f_i} + -\frac{\partial(\overline{u_i u_j})}{\partial x_j} +\frac{\partial(\bar{u}_i \bar{u}_j)}{\partial x_j} \\
\frac{\partial \bar{u}_i}{\partial t} + \frac{\partial(\bar{u}_i \bar{u}_j)}{\partial x_j} = -\frac{\partial \bar{p}}{\partial x_i} + \nu \frac{\partial^2 \bar{u}_i}{\partial x_j \partial x_j} + \bar{f_i}-
\underbrace{\frac{\partial \tau_{ij}}{\partial x_j}}_{\text{subgrid stess tensor}}
\end{equation}

In LES small scales are not explicitly resolved. Any scales that are smaller than the filter width are referred to as sub-grid-scales (SGS).
Thus the SGS term has to be modeled. As in three-dimensional Navier-Stokes turbulence the nonlinear interactions of turbulent velocity fluctuations to leading order give rise to a spectral flux of energy towards smaller scales, the main purpose of a SGS-model in such a system is to generate the right amount of energy dissipation with a spatial 
stress distribution that is consistent with the one due to the original velocity field.
As the original velocity field is unknown to the SGS-model, this closure problem can only be fulfilled up to a certain level of approximation that is dependent on
particular physical focus and goals of the LES modeler. Thus, many sub-grid models of differing complexity and physical background have been proposed, see e.g. \cite{meneveau2000scale-invariance,biferale2019selfsimsgs,mueller2002dynamic,Meneveau2010SGS} and references therein.

%The best known SGS-model is the Smagorinsky-Lilly model~\cite{smagorinsky1963,lilly1967representation}, which yields realistic values for locally isotropic turbulence. Other eddy viscosity models~\cite{deardorff1973use,schumann1975subgrid} have improved modeling of time development of small scales. Dynamic SGS models~\cite{germano1991,lilly1992germanoSGS,meneveau1996,ghosal1995dynamic} use the Germano identity~\cite{germano1992turbulence}, which describes how model parameters for the unresolved scales may be adapted from the resolved scales in a self-consistent manner under the assumption of approximate self-similarity of the small-scale fluctutations. Other types of models~\cite{clark1979evaluation,bardina1980improved,liu1994properties} rely on the deconvolution method~\cite{stolz1999approximate}, that assumes an inverse operation to the LES filter to recover an approximation to the original velocity field.

\subsection{Related Work}
In recent years effort to apply machine learning to the field of fluid dynamics has been made~\cite{brunton2020machine, chang2019MLframeworks, tompson2017accelerating}. As it is often intractable to resolve fine grained features of the solution, many approaches incorporate machine learning tools to facilitate computation of the coarse grained representation. To achieve this, many approaches use high resolution data from DNS as ground truth that is then downsampled. To incorporate physical constraints, the ML model is usually restricted to reconstructing only a part of the unknown solution while other parts are described using the exact dynamics or more traditional approximations~\cite{bar-sinai2019discretizations}. In the case of turbulence, most approaches focus on learning the SGS closure end to end by estimating the difference between downsampled DNS and LES data using an artificial neural network~\cite{gamahara2017searching, beck2019deep, lapeyre2019training, pawar2020priori, rosofsky2020artificial, maulik2019subgrid, bode2019deep}. While the trained networks are initially dissipative when applied in a new setting they can lack long-term stability as high frequency errors accumulate. To counteract these drawbacks effort has been made to incorporate physical principles into the machine learning models~\cite{wang2020towards, taghizadeh2020turbulence, king2018deep, Kochkov2021MLaccelerated}. 

In the present paper, we follow an alternative strategy where the model is non-linearly adapting to the flow that develops under its influence. This means that, unlike previous work, we formulate our objective without referring to the underlying `velocities' as targets but instead aim to tune the parameters to reproduce statistical characteristics of the system.

Our work is most closely related to recent work on incorporating machine learning models into differential equations. Specifically the approach to approximate not the quantity of interest itself but its derivative with a machine learning model. This differential equation can then be solved using standard numerical methods. 
To optimize the model's parameters $\theta$ a loss function $L$ needs to be defined. This loss is usually a scalar quantity that either directly compares the calculated solution of the differential equation with ground truth data, or minimizes some other quantity that is dependent on the solution. In order to update the model's parameters so that they may minimize the loss one needs to calculate the derivative of the loss with respect to the parameters $\frac{\d L}{\d \theta}$. In standard machine learning procedures this is usually done using automatic differentiation. In the present case this approach can become very memory inefficient as one needs to differentiate through the steps made by the numerical solver. Thus the gradient $\frac{\d L}{\d \theta}$ is approximated by solving an additional differential equation, the adjoint, backwards in time.  This idea was popularized recently by~\cite{Chen2018NeuralODE}, although previously discussed in~\cite{lecun1988theoretical} and~\cite{pearlmutter1995gradient}. We recap this approach briefly in Section~\ref{sec:adjoint}.

%Most of the above-mentioned work follows the standard ML-paradigm of model identification and training based on a predefined training set of `true' turbulence. This approach has shown some success but, e.g.,  has large resource requirements regarding memory and computing time while being rather inflexible with regard to changes in the flow configuration. Taking the important temporal correlation of the velocity field into account renders such an approach even more cumbersome. 

The present paper focuses on the basic properties of the problem and the  potential pitfalls that our
learning-by-evolution approach entails. We therefore  attempt to reduce mathematical complexity by replacing the full Navier-Stokes equations by a much simpler shell model. 

\subsection{GOY model}
The Gledzer-Ohkitani-Yamada (GOY) model~\cite{Gledzer, OhkitaniYamada} is a differential equation given by
\begin{align}
\frac{d}{dt} u_i &= -\underbrace{D_i}_{\text{viscous dissipation}} + \underbrace{F_i}_{\text{forcing}} + \underbrace{\i C_i}_{\text{quadratic interactions}} \label{eq:GOY}\\ \nonumber\\
D_i &=\nu k^2_i u_i  \label{eq:dissipation}\\
F_i &=f\delta_{i, j} \label{eq:forcing}\\
C_i &= k_i u^*_{i+1}u^*_{i+2} - \epsilon k_{i-1} u^*_{i-1}u^*_{i+1} + (\epsilon-1) k_{i-2}u^*_{i-1}u^*_{i-2}\label{eq:non_linear}
\end{align}
where $u$ is the complex velocity, $u^*$ denotes the complex conjugate, $\i$ is the imaginary unit, $\nu, k_0, \lambda, \epsilon$ are real parameters, and $f$ is a complex parameter. The so called shell numbers are denoted by $k_i = k_0\lambda^i$, $i=1, 2, ... , N$.

\begin{figure}
\centering
\includegraphics[width=0.9\linewidth]{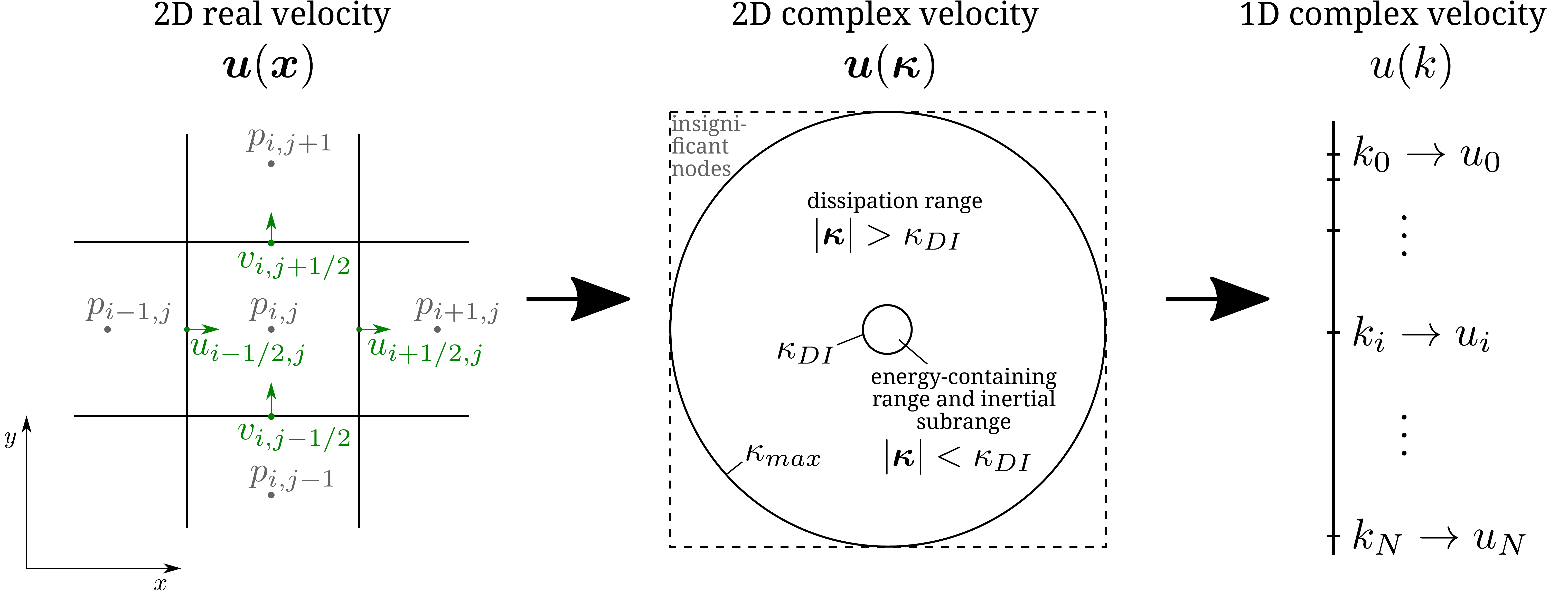}
 \caption{Graphical intuition for the GOY model. Left: computational grid for 2D velocity components in real space~\cite{bridson2007fluid}, center: solution domain in wavenumber/Fourier space~\cite{pope2001turbulent}, right: exponentially spaced wavenumbers with scalar complex scalar velocities}
\label{fig:toGOY}
\end{figure}

A basic source of inspiration for this model are the Navier Stokes equations in Fourier space
\begin{align}
\frac{\partial}{\partial t} \F\lbrace u_i\rbrace &= -\nu |\boldsymbol{\kappa}|^2\F\lbrace u_i\rbrace + \F\lbrace f_i \rbrace -\i \kappa_m\F\lbrace u_i u_m\rbrace + \i \kappa_i \frac{\kappa_l \kappa_m}{|\boldsymbol{\kappa}|^2}\F\lbrace u_l u_m\rbrace 
\label{eq:NSE_fourier}
\end{align}
where instead of having a real valued velocity that depends on spacial coordinates we have a complex valued velocity depending on wavevectors $\boldsymbol{\kappa}$. The operator $\F$ denotes the Fourier-transformation. Large scales, $\ell$, of turbulent motion correspond to Fourier modes with small wavenumber $\kappa=|\boldsymbol{\kappa}|$. The simplification of the GOY model then lies in reducing all velocity vectors of equal wavenumber $|\boldsymbol{\kappa}|$ to a scalar complex velocity $u_i$, that depends on the scalar wavevector $k_i$ (an intuition is given in Figure~\ref{fig:toGOY}). The $k_i$'s are also called shell numbers and are usually chosen to be geometrically spaced to cover a wide range of turbulent scales. Shell models, like the GOY model, have been studied extensively as they can model the statistical scaling properties of turbulence at high Reynolds numbers while keeping computational cost low~\cite{ditlevsen_2010, Biferale}.

Throughout this work we use standard parameters
\begin{align*}
\epsilon &= 0.5 \\
\lambda &= 2\\
\nu &= 10^{-8}\\
k_0 &= 2^{-4}\\
F_i &= f\delta_{i,4}= \left\{\begin{matrix}
5\times 10^{-3}(1+\i) & \text{for $i=4$}\\
0.0 & \text{else}
\end{matrix}\right.\\
u^{(0)}_i &= \left\{\begin{matrix}
1\times10^{-5}(1+\i) & \text{for $i \in \{3, 5\}$}\\
0.0&\text{else}
\end{matrix}\right.\\
N &= 22
\end{align*}

\begin{figure}
\centering
\includegraphics[height=0.3\linewidth]{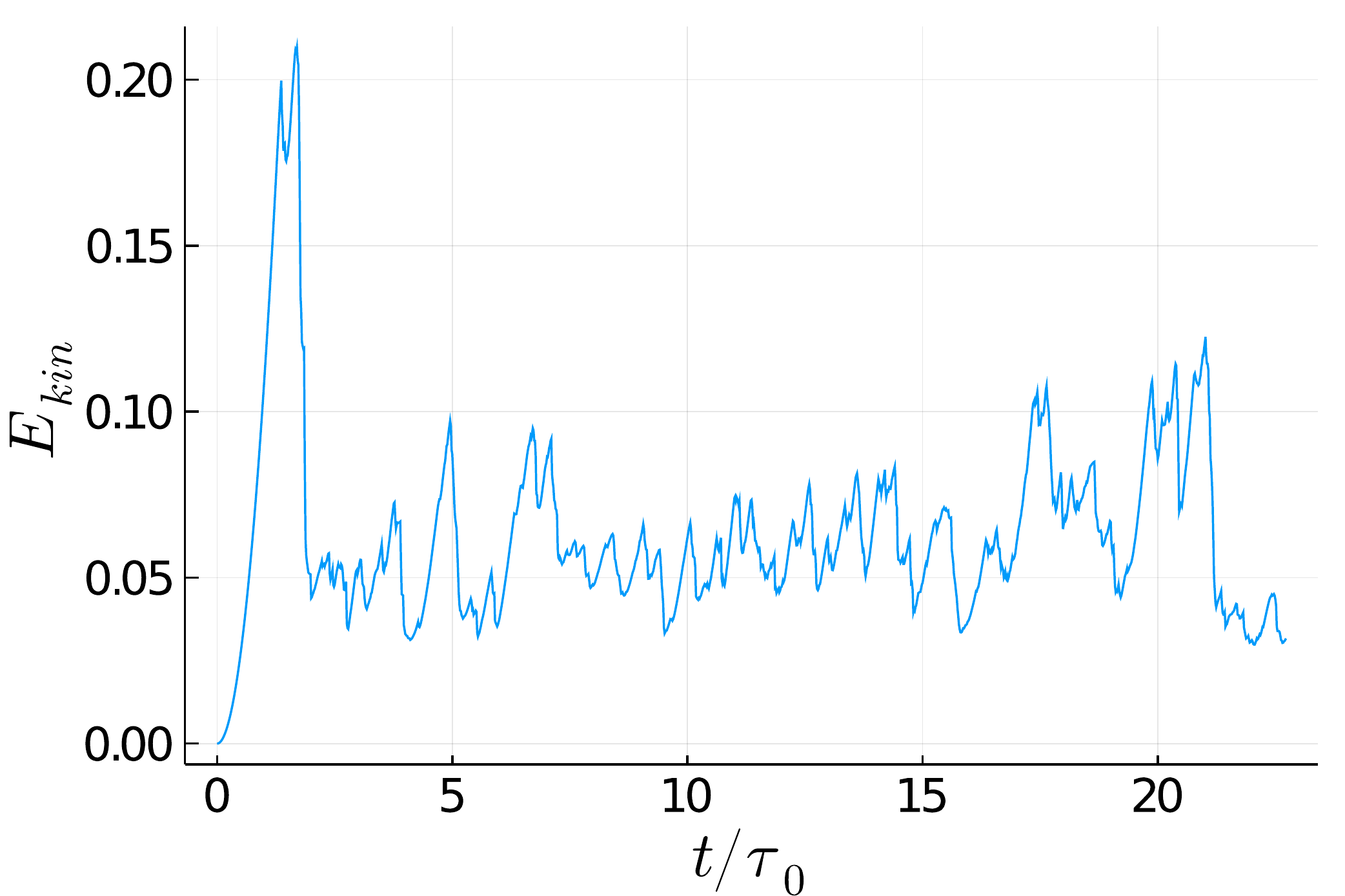}
\hskip 0.5cm
\includegraphics[height=0.3\linewidth]{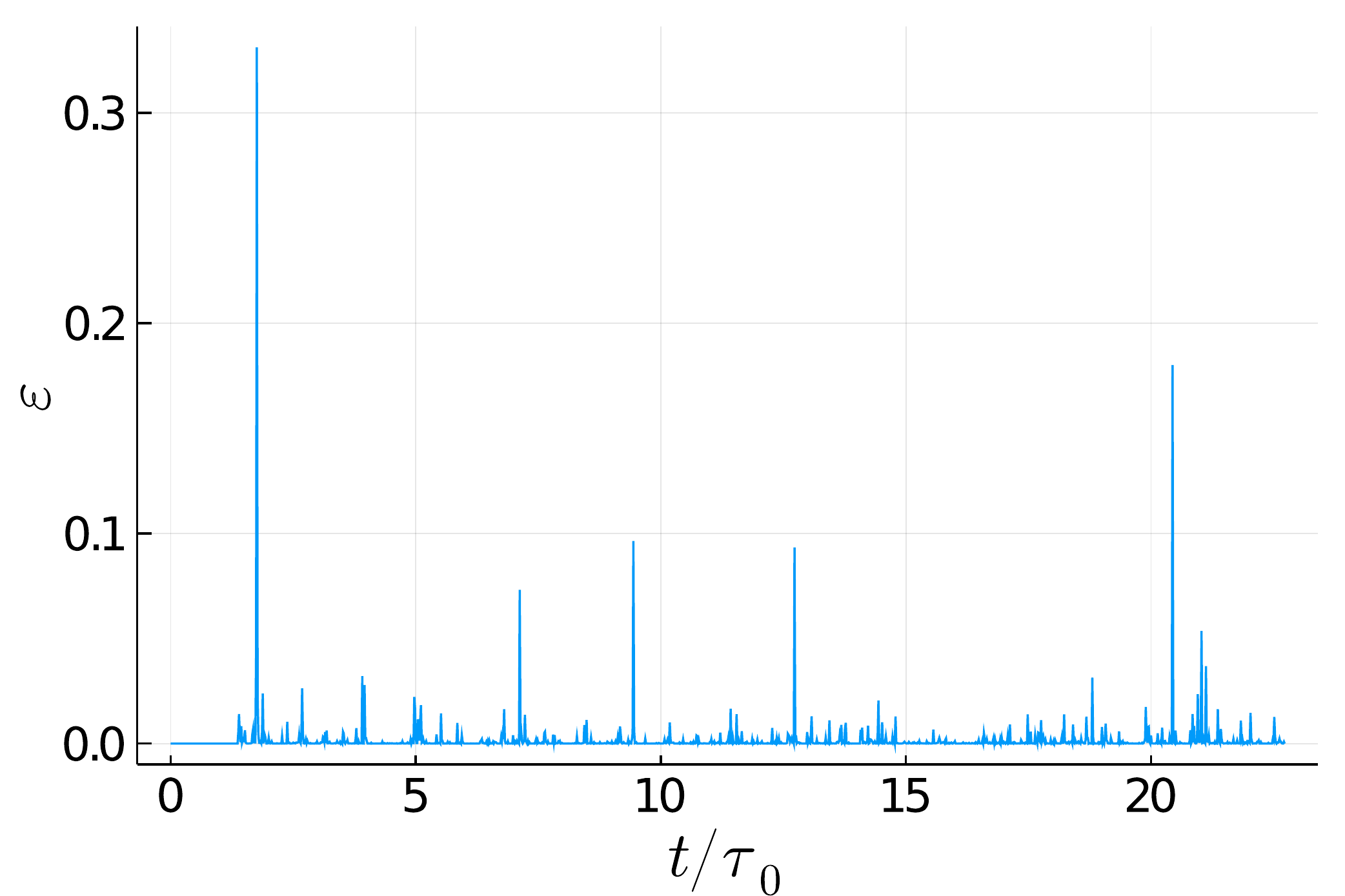}
\caption{After an initial period where energy accumulates in the system, interactions between neighbouring components eventually transport energy to larger shells, where it is dissipated, on average at the rate of which new energy is introduced into the system via forcing. Left: total kinetic energy in the system. Right: dissipation rate}
\label{fig:GOY_simulation}
\end{figure}

We show an example of a simulation of 1500 time steps (around 23 large eddy turnover times) with the above parameters in Figure~\ref{fig:GOY_simulation}. The left graph shows total kinetic energy $E_{kin} = \sum_{i=1}^N \frac{1}{2} |u_i|^2$ and the right graph shows the dissipation rate $\epsilon = \sum_{i=1}^N \nu k^2 |u_i|^2$

We plot the quantities over turnover times of the largest fluctuation $\tau_0 = \left< k_1|u_1|\right>^{-1}$, where $\left<\bullet \right>$ denotes the average over time. 
The nonlinear turnover time estimates how long it takes for a fluctuation to be reduced to Kolmogorov scale. Averaging over at least 10 large eddy turnover times is relevant for statistical quantities.

\subsection{Statistical quantities}\label{sec:statistical_quantities}

\begin{figure}
\centering
\includegraphics[height=0.4\linewidth]{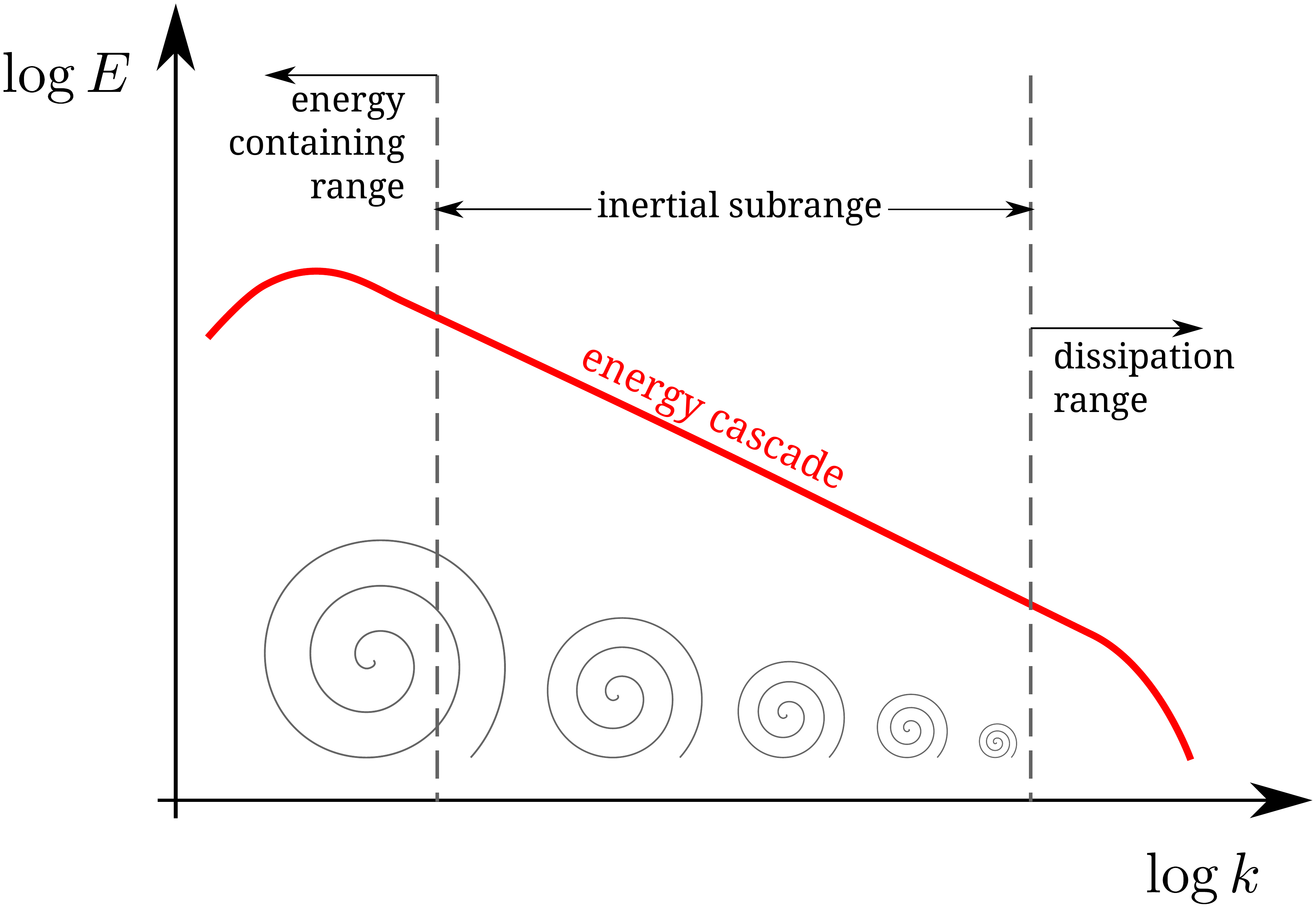}
\caption{Energy cascade involving the transfer of energy from large scales to small scales of motion.}
\label{fig:energy_cascade}
\end{figure}

In turbulent flows we can observe the non-linear interaction dynamics of turbulent fluctuations on a broad range of spatial and temporal scales of motion. This gives rise to efficient mixing of quantities advected by the flow and the roughly self-similar destruction and emergence of structural coherence on spatial scales of different size (break-up or merging). Although the structural dynamics appears to be random, it results on average in directed and conservative spectral transport of ideal invariants, such as energy or helicity, termed cascade. 
The cascade direction, either towards larger or smaller wavenumbers,  is dependent on the type of quantity, the underlying physics, the dimensionality and the general configuration of the system under consideration. 
Stationarity of turbulence is thus only possible if appropriate sources and sinks of the cascading quantities are present at largest and smallest scales of the flow.
Three-dimensional hydrodynamic Navier-Stokes turbulence exhibits a direct cascade of kinetic energy per unit mass. The cascade generates an energy flux from large scales (small wavenumbers) where kinetic energy is injected by some physical process to small scales (large wavenumbers) where kinetic energy is removed by dissipation, see Figure~\ref{fig:energy_cascade}. 

Although the GOY model is a very simplified version of the Navier-Stokes equations it produces statistical similarity properties that are comparable to those of real world turbulence. These statistical characteristics  are found in the inertial subrange of spatial scales, where neither external forcing, nor dissipation dominates and where energy transfer happens entirely by spectrally local, nonlinear interactions between fluctuations at different but similar scales of motion.

\begin{figure}
\centering
\includegraphics[height=0.3\linewidth]{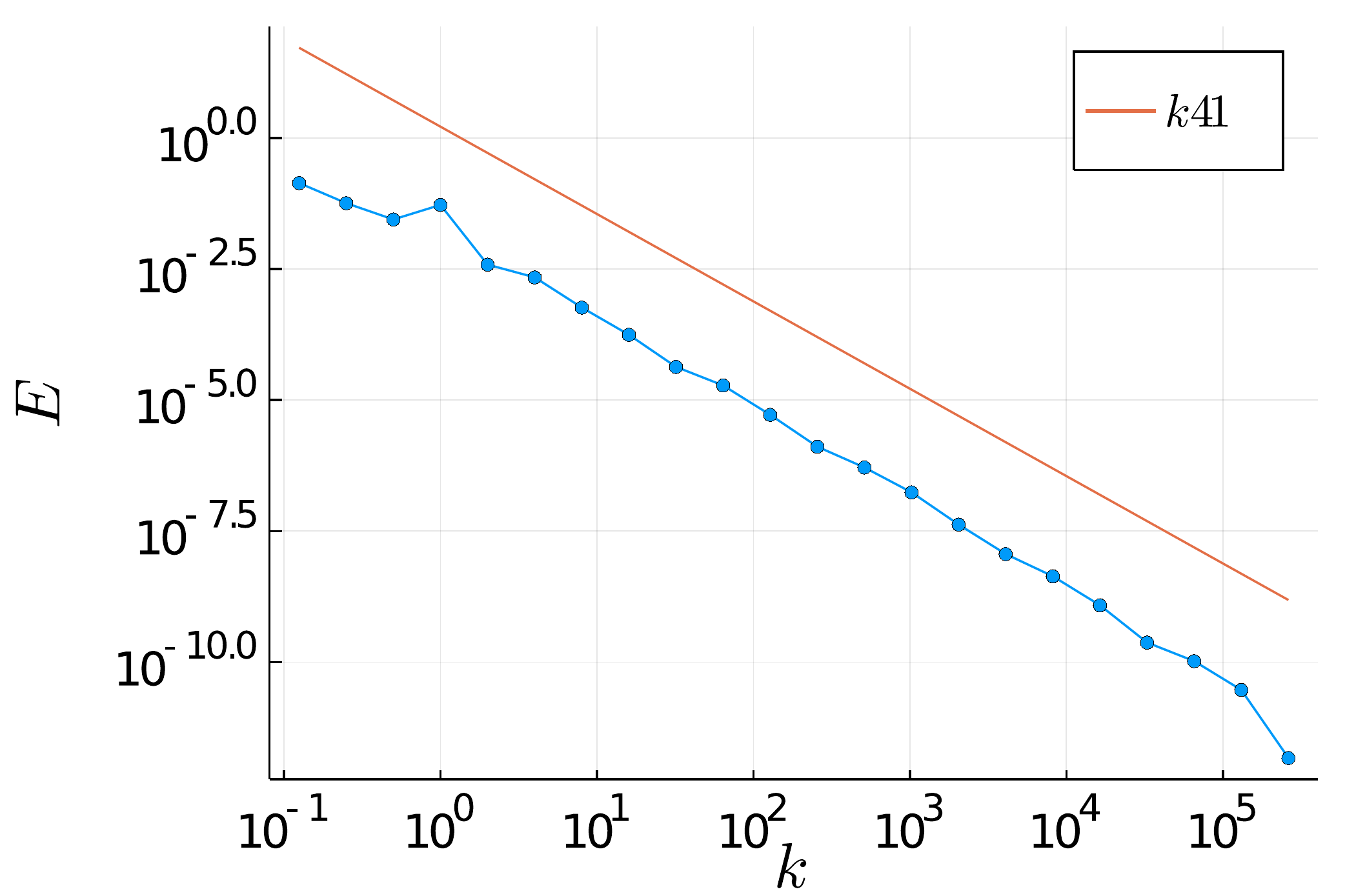}
\hskip 0.5cm
\includegraphics[height=0.3\linewidth]{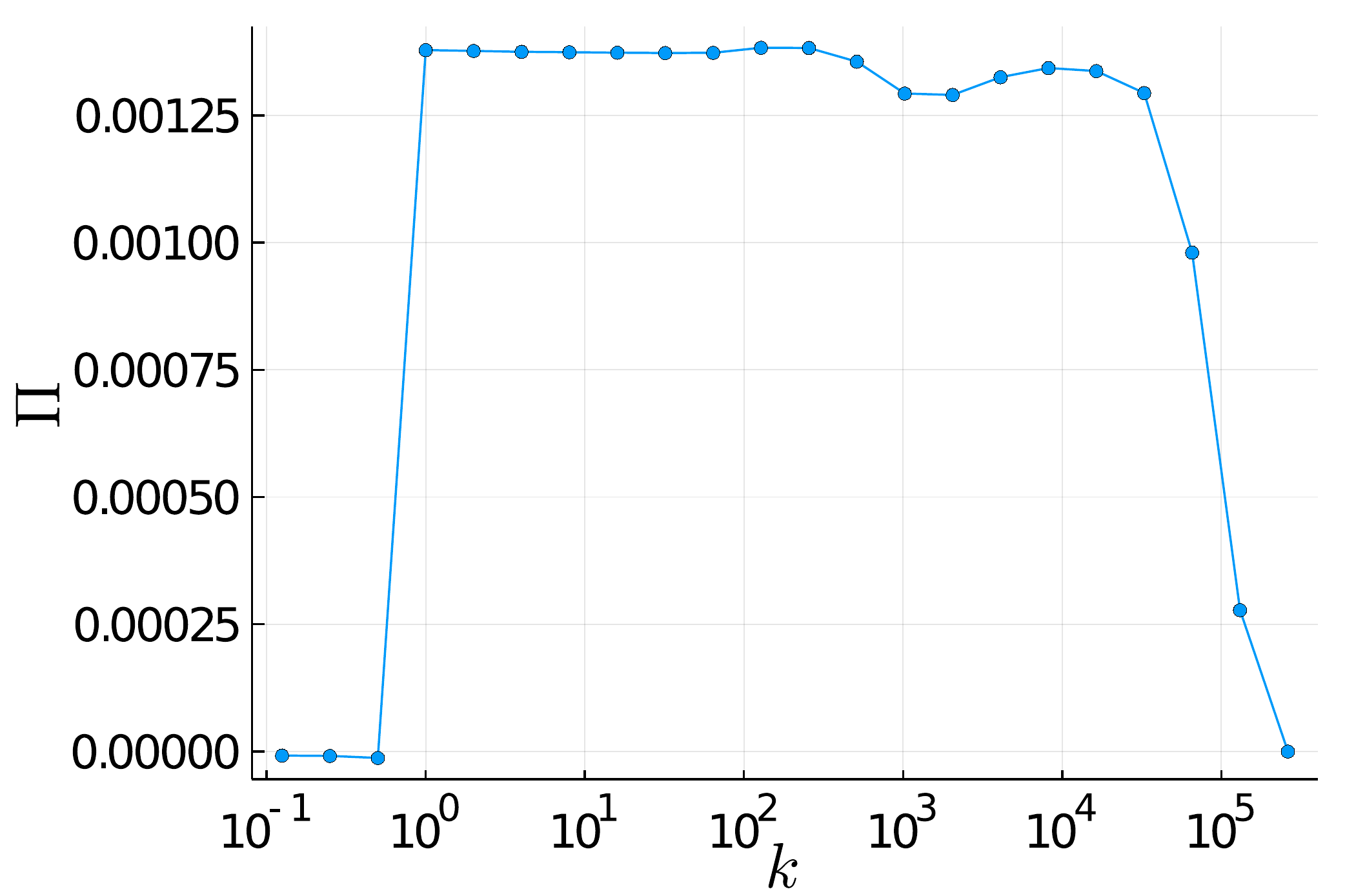}
\caption{Left: energy cascade with characteristic $-\frac{5}{3}$ slope within inertial range, Right: non linear energy flux within the inertial range}
\label{fig:GOY_cascade_flux}
\end{figure}

The energy density distribution over the shells
\begin{equation}
E_i =  \frac{1}{2}\frac{|u_i|^2}{k_i}
\label{eq:energy_density}
\end{equation}
approximates the characteristic slope of the Kolmogorov $-\frac{5}{3}$ spectrum~\cite{frisch1995turbulence, kolmogorov1991k41} in the inertial range $2^2 \leq k_i \leq 2^{14}$, see Figure~\ref{fig:GOY_cascade_flux}.

Another statistical quantity of interest is the non-linear energy flux (see Figure~\ref{fig:GOY_cascade_flux}) which is determined by the non-linear interactions between the shells described by equation~(\ref{eq:non_linear}):
\begin{equation}
\Pi_i = -\Im\{u_{i}u_{i+1} (k_i u_{i+2}+(1-\epsilon)k_{i-1}u_{i-1})\}
\end{equation}
where $\Im\{\bullet\}$ denotes the imaginary part of the argument. In the case of Kolmogorov scaling of the energy spectrum, the energy flux in the inertial range is roughly constant as energy is transported conservatively from lower to higher shell numbers.

%Figure~\ref{fig:GOY_cascade_flux} shows the two quantities over the shell numbers.

\subsection{Our Modification}
To create a learning task we drop the term describing the dissipation and aim to relearn this term.
We use the original GOY equations~(\ref{eq:GOY}) for reference to the ground truth. Our modified GOY equation is then

\begin{equation}
\frac{d}{dt} u_i = \underbrace{F_i}_{\text{forcing}} + \underbrace{\i C_i}_{\text{quadratic interactions}} -\underbrace{M _i}_{\text{machine learning model}}\,.
\label{eq:GOY_modified}
\end{equation}

The term $M_i$ can be dependent on $u, t$ and parameters $\theta$. 
We define our model as 
\begin{equation}
M_i=\theta k_i^2 u_i
\label{eq:ml_model}
\end{equation}
with learnable parameter $\theta$. This simple model mimics a diffusive Laplacian dissipation term
with $\theta$ as the adjustable dissipation coefficient.
\section{Theoretical considerations}
In this paper, the optimal solution is characterized by a target value 
of a certain physical characteristic of the velocity field, i.e. the asymptotic self-similar scaling exponent of the energy spectrum which is a statistical two-point correlator. This quantity acts as a statistical sensor providing information on the model's performance.  

The proposed learning-by-evolution technique is conceptually closer to common strategies to understand physical behaviour than the classical ML perspective that regards a physical system as a set of typically sparse but correlated data. Our approach is 
able to straightforwardly include physical, in particular statistical properties
of the system under consideration which constrain the ML-algorithm.
While the region determining spatial statistics in homogeneous turbulence is by definition unbounded, this is generally not the case for temporal statistics of a system regulated by some feedback control mechanism. The combination of an SGS-model and the attached ML-algorithm which evaluates the reaction of the system on the model via a statistical sensor is fundamentally constrained by three timescales:
i) the propagation timescale $\tau_\text{p}$ characterizing the time required to communicate a change of state of the model throughout the parts of the system that determine the state of the statistical sensor,
ii) the relaxation timescale $\tau_\text{r}$ on which the statistical sensor converges within a defined variation $\delta_m$ towards a new stationary state after a modification of the SGS-model has been communicated to the relevant parts of the system, and iii) the timescale of backwards-stationarity, $\tau_\text{b}$, which determines the time-horizon up to which the ML-algorithm can inspect
the past evolution of the system to optimize the model's parameters. 
Evidently, for the interval $\tau_\Delta$ between model modifications, we have to require $\tau_\Delta\ge\tau_\text{p}+\tau_\text{r}$.
It is important to note that the above relation represents
a fundamental physical constraint, since propagation and relaxation timescales are generally independent of the specific technicalities of the model or the ML-implementation.
Changes to the model that occur faster than $(\tau_\Delta)^{-1}$
cannot be distinguished by the statistical sensor and do not allow to control the model in a well-defined manner.  
While the propagation timescale represents a strict lower bound on $\tau_\Delta$, the relaxation time of the sensor is mainly determined by the required level of statistical convergence and the tolerable deviation from forward-stationarity, i.e. statistical stationarity reached \textit{after} a modification of the model.  
Therefore, $\tau_\text{r}$ should be significantly  longer than the longest auto-correlation timescale of fluctuations determining the sensor state.
Both timescales can be estimated by physical considerations:
a relevant change of state of the SGS-model will lead to a different total dissipation rate of energy, $\varepsilon$. For decaying turbulence, $\varepsilon=-\mathrm{d}/\mathrm{d}t E(t)$.
For driven turbulence that has attained a statistically stationary state this rate 
has to be equal to the rate of energy injection by the driving mechanism which is usually known in numerical experiments.
Since the turbulent energy cascade towards small-scale is conservative any statistically significant variation of $\varepsilon$ will backpropagate from the smallest towards the largest scales  of the flow on the order of a large-eddy turnover time $\tau_0$.
%, which can be approximated dimensionally as $\tau_0\sim E/\varepsilon$. 
The auto-correlation timescale characterizing the relaxation interval of the sensor is roughly estimated to be of the order of $\tau_0$ as well since in turbulence DNS or LES scale separation of inertial range and large-scale driving is often severely limited by finite numerical resources.
Consequently, the temporal evolution of the system as observed by the ML-algorithm has a granularity of the order of $\tau_0\sim\tau_\text{p}+\tau_\text{r}$.

Another important aspect that depends on physical properties of the turbulent flow is
the maximum time-horizon limiting the extent to which the ML machinery can use data from the past to optimize the present model configuration.
This introduces a third timescale, $\tau_\text{b}$, on which statistical measurements of the current state of the flow are permissible, i.e. on which the system's state can be assumed as statistically stationary. Causes for deviations from backwards-stationarity, i.e. statistical stationarity \textit{before} a model modification, are for example a secular drift of physical parameters or, more importantly, a preceding modification of the model  by the ML-algorithm. The points in time characterizing such changes of the system are strictly speaking boundaries between statistically non-equivalent states that should
not be combined to evaluate the model's performance. 
In summary, rather basic considerations
require a finite minimum interval, $\tau_\Delta$  between subsequent model modifications as well as a finite time horizon for inferring the model's performance from the preceding evolution of the system.

This reasoning, however, does not necessarily lead to an efficient optimization of the model if the required path of model modifications allows states of the system for which the evolution is difficult or even impossible to compute. This is the case in the present work where we intend to find a model for a diffusive dissipation term, $\nu \Delta u$. The exponential behaviour associated with it can give rise
to growing unstable solutions for $\nu<0$ or exponentially fast decay. A choice of $\nu$ that is either 
positive, but too large, or negative leads to numerical stiffness or even complete loss of a stationary solution. As a consequence, the ML-algorithm has to react to its own modifications
of the model on much shorter timescales than estimated above to avoid being deadlocked in a state of the system that prevents further numerical evolution and improvement. 
We define the nonlinear turnover time on spatial scale $\ell$ as $\tau_\text{NL}=\ell/u_\ell$ with a characteristic root-mean-square velocity fluctuation $u_\ell$ on that scale.
For a spectral energy scaling $E(k)\sim k^{-\gamma}$ with $k\sim\ell^{-1}$ and $u^2\sim kE(k)$, we obtain $\tau_\text{NL}\sim (k^3 E(k))^{-1/2} \sim k^{(\gamma-3)/2}$. Thus, the turnover time in the inertial range of Kolmogorov turbulence decreases with scale as a power-law and the system, in particular the numerically permissible timestep, 
responds much faster to modifications of the model than assumed above. 
Consequently, in order to prevent a numerically problematic evolution, a sub-optimal choice of model parameters has to be detected on the shortest timescales related to the statistical sensor.
In the present work the model has to be re-adjusted almost every timestep which is determined by the turnover
time on the scale of the numerical grid.

\subsection{Optimization Problem}

To train the model we need to define an objective function that is being minimized by retrospective
analysis of the systems's immediate past.
We want the scaling exponent of the energy shell spectrum  to resemble the slope given by Kolmogorov's theory~\cite{frisch1995turbulence}. For that we take the mean squared error of the logarithmic slope at every shell and the constant $-\frac{5}{3}$. The loss is then given by

\begin{equation}
L = 1/N\sum_i \left(\frac{\log(E_{i+1})-\log(E_i)}{\log(k_{i+1})-\log(k_i)} - \frac{5}{3}\right)^2
\label{eq:objective}
\end{equation}

where $E_i$ denotes the energy density from equation~(\ref{eq:energy_density}) at shell number $k_i$.

\subsection{Adjoint method}\label{sec:adjoint}

Our ML model is part of the differential equation. When we solve the differential equation to get the velocity value at $t_1$, we integrate the differential equation using a numerical solver like the Runge-Kutta Order 4 method with adaptive step width.
The problem can be formally expressed as:
\begin{align}
u(t_1) &= u(t_0) + \int_{t_0}^{t_1} \frac{\d u(t)}{\d t} \d t\\
 &= u(t_0) + \int_{t_0}^{t_1} F + \i C(u) - M(u, t, \theta) \d t
\end{align}

To update the parameters $\theta$ of the model we need the gradient of the loss function, given by equation~(\ref{eq:objective}), with respect to our parameters $\frac{\d L(u(t))}{\d \theta}$. In traditional ML problems this gradient is calculated with standard backpropagation using automatic differentiation~\cite{rumelhart1986backprop, GoodfellowBook}. For that, intermediate results in the forward pass are usually saved. In the case at hand we would have to backpropagate through all the solver steps that were taken in the forward pass, which can be very memory inefficient. An efficient method to calculate gradients is the finite difference method that only requires access to the loss itself $L(u(t))$ but this can become unfeasible if our model has many parameters since we need to calculate a complete forward pass per parameter. However there is another method to calculate the gradients approximately: the adjoint sensitivity or adjoint state method. It has many applications in physics and more recently in neural networks~\cite{Chen2018NeuralODE}. The idea is to approximate the desired derivative of the loss with respect to the parameters $\frac{\d L}{\d \theta}$ by solving an integral numerically:
\begin{equation}
\frac{\d L}{\d \theta} = -\int_{t_1}^{t_0} \lambda(t)^T \frac{\partial f}{\partial \theta}
\end{equation}

where $f = \frac{\d u}{\d t}$.
To this end the so called adjoint is introduced as the derivative of the loss with respect to a specific state
\begin{equation}
\lambda(t) = \frac{\partial L}{\partial u(t)}
\end{equation}

The adjoint is the solution to the initial value problem
\begin{equation}
\dot{\lambda}(t) = -\lambda^T \frac{\partial f}{\partial u}\quad\text{subject to}\quad
\lambda(t_1) = \frac{\partial L}{\partial u(t_1)}
\end{equation}

The two integrals are solved backwards in time. The forward pass, that is the integration forward in time to solve for velocity $u(t_1)$, can be treated as a black box. We only need to know how our differential $f=\frac{\d u}{\d t}$ depends on $u$ and parameters $\theta$ and how our loss depends on the final state $\frac{\partial L}{\partial u(t_1)}$. For more details see \ref{appndx:adjoint}.

\subsection{Technical requirements}
Handling of complex numbers and implementation of adjoint methods and numerical solvers are rarely required for standard ML tasks, thus the most common Python-based libraries for machine learning, like PyTorch or Tensorflow, currently lack the proper tools for our problem. This is why we chose Julia-based libraries for implementation. They offer a vast selection of numerical solvers, implementations of adjoint gradients, and have good support for complex numbers. For the learning algorithm we use Adam~\cite{kingma2015adam} as an optimizer as it performs well on problems with noisy gradients~\cite{Ruder2016AnOO}. The learning rate, $\alpha$, scales the updates to the parameters $\theta$ during retrospective training, and essentially sets the pace at which the ML-algorithm is moving in parameter space. 
As the adjusted model parameter has a direct impact on the maximum timestep allowed for numerical integration, the learning rate should be sufficiently small to allow for automated corrections if model performance becomes worse. 
As the loss is technically an average of shell energies, two physical timescales are
relevant in this respect: i) the nonlinear turnover time $\tau_\text{NL}(k_i)$ characterizing nonlinear spectral energy transfer, and ii) the model timescale $\tau_\text{M}(k_i)$ which for the diffusive model \eqref{eq:ml_model} is estimated at wavenumber $k_i$ as $\tau_\text{M}=(2k_i^2\theta)^{-1}$. For $i=22$ the latter relation links the largest, numerically permissible timestep, $\Delta t$, with the model parameter $\theta$ as $\theta \Delta t\sim 10^{-11}$.  Thus, for the given GOY-parameters a reasonable choice for the learning rate amounts to $\Delta \theta=1\cdot 10^{-9}$, while all other hyperparameters of the optimization algorithm are set to standard values. The Adam algorithm effectively restricts the absolute values of the updates to the learning rate; $|\Delta \theta| \lessapprox \alpha$. Large values for $\alpha$ can thus lead to problems with unstable solutions and stiffness of the differential equation (see Section~\ref{sec:failure_modes}).

\section{Experiments}

\subsection{Setting the dissipation to zero}
\begin{figure} 
\centering
\includegraphics[height=0.21\linewidth]{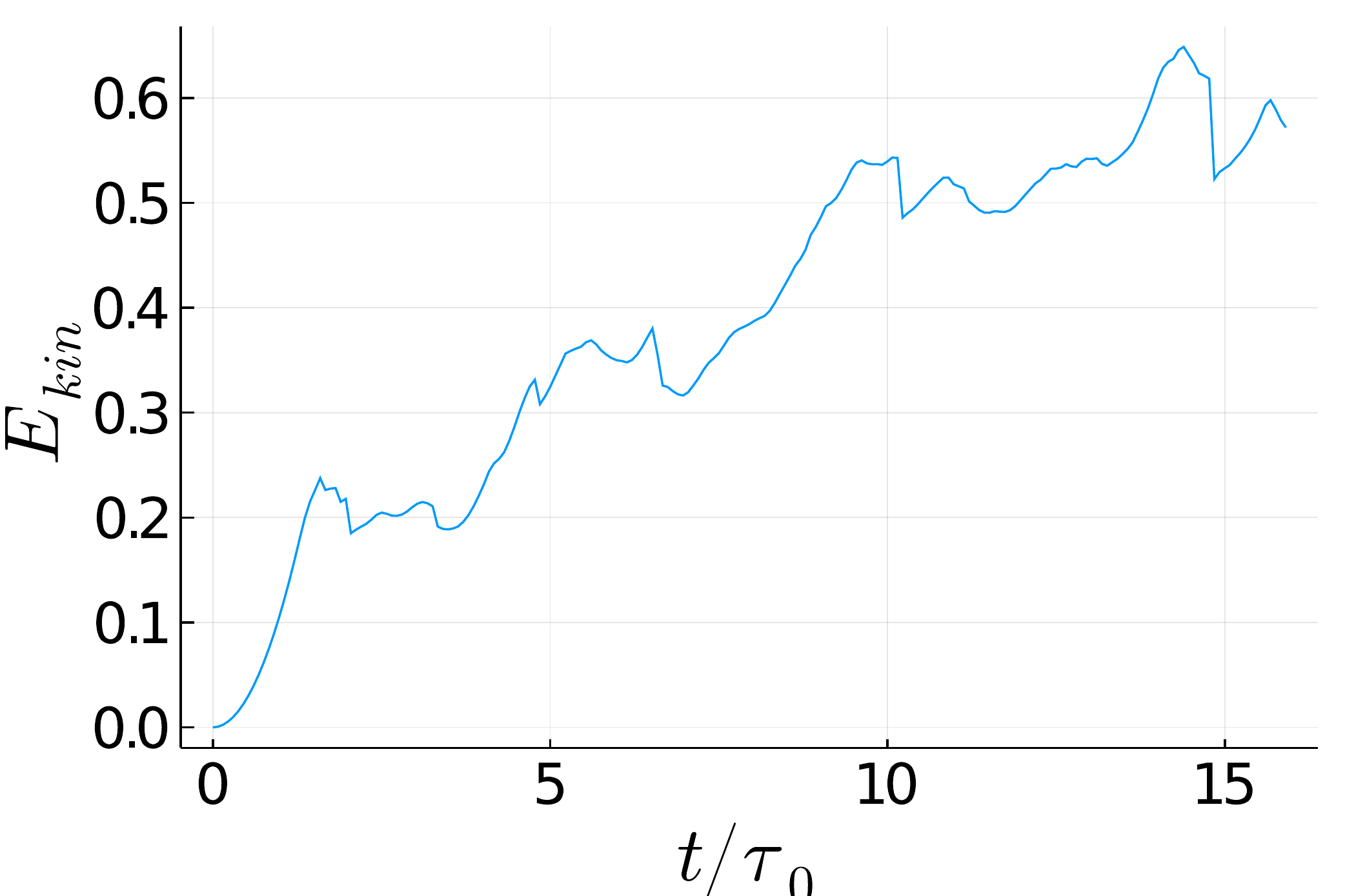}\hskip 0.1cm
\includegraphics[height=0.21\linewidth]{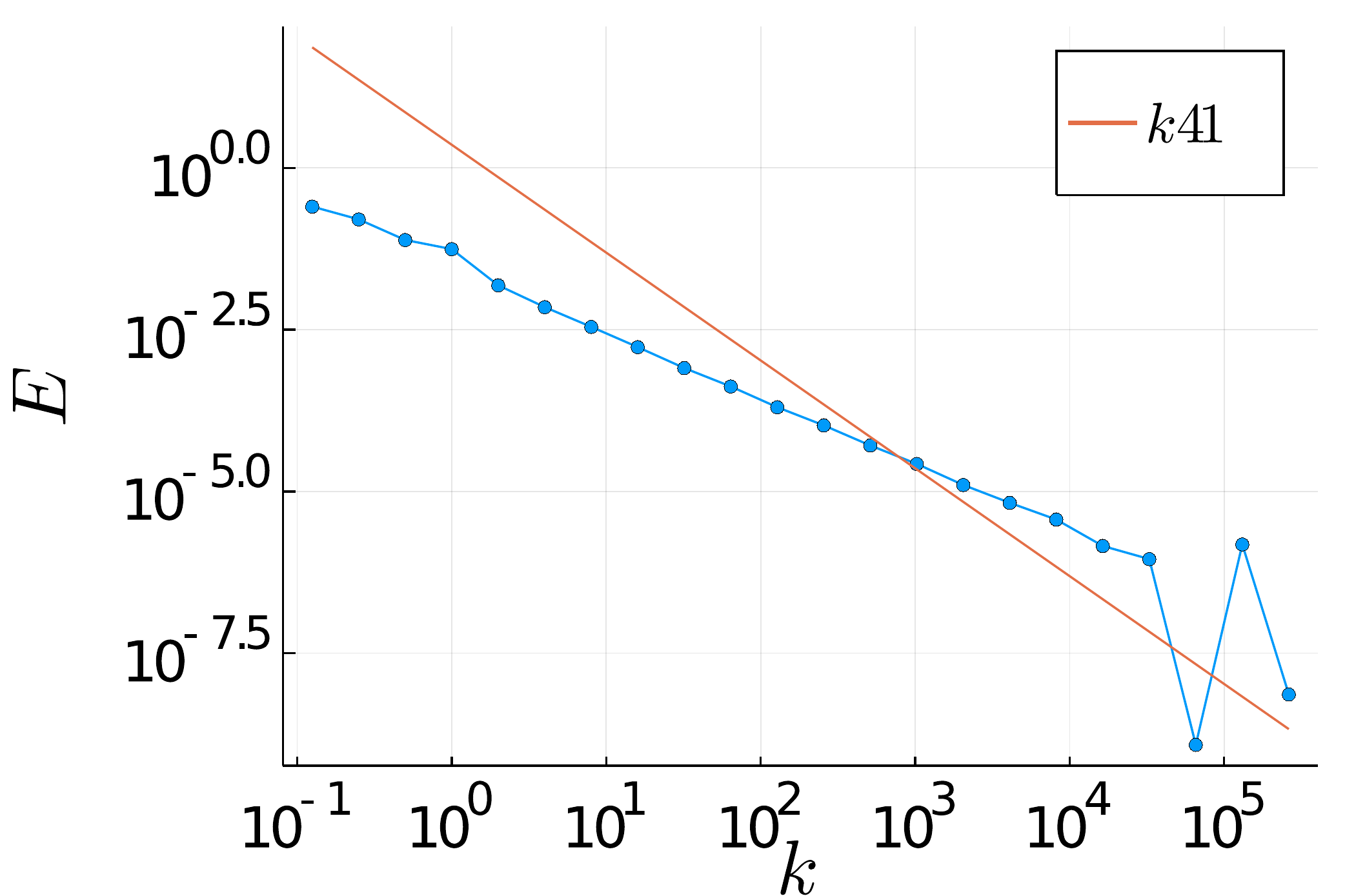}
\hskip 0.1cm
\includegraphics[height=0.21\linewidth]{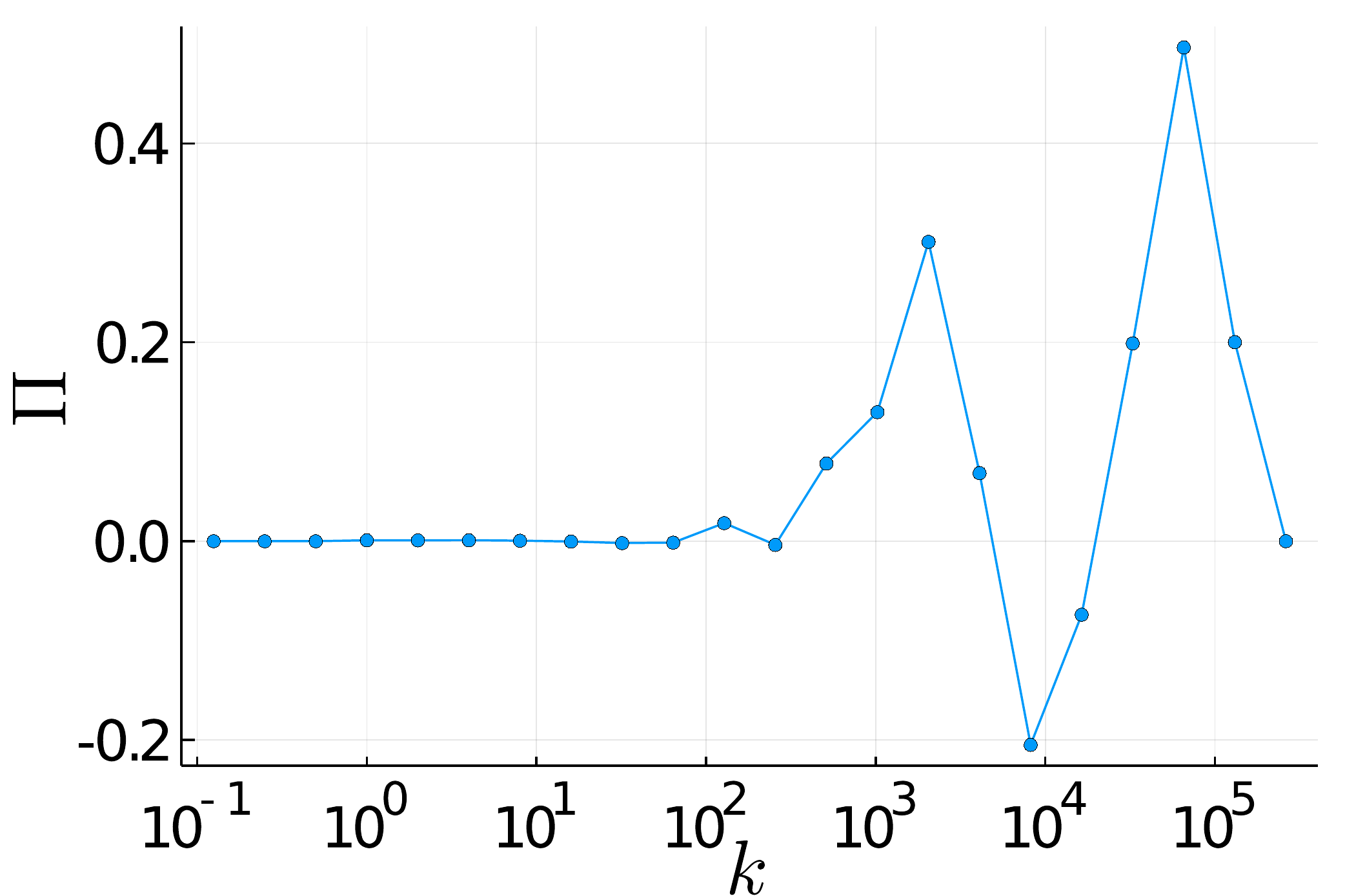}
\caption{Simulation results without dissipation term. Left: accumulation of total kinetic energy; Middle: energy scaling deviates from characteristic $-\frac{5}{3}$ value; Right: non linear energy flux}
\label{fig:GOY_simulation_nu=0}
\end{figure}

We already showed simulation results for a GOY system with standard parameters in Figure~\ref{fig:GOY_simulation} and~\ref{fig:GOY_cascade_flux}. For the modified equations we have replaced the dissipation term with a learnable model. If we set this model to zero, and repeat the simulation, energy flux and energy spectrum divert from the Kolmogorov behaviour. Due to the direct energy cascade towards small-scales generated by the GOY model and the lack of small-scale dissipation, energy piles up at smallest scales.
 Figure~\ref{fig:GOY_simulation_nu=0} shows simulation results when the dissipation term is removed and no alternative model is introduced. Energy accumulates in the system, the energy spectrum looses its characteristic inertial-range slope of $-\frac{5}{3}$ and the non-linear energy flux is no longer even roughly constant in the scaling range.

\subsection{Learning $\nu$}
\label{sec:learning_nu}

\begin{figure}
\centering
\includegraphics[height=0.30\linewidth]{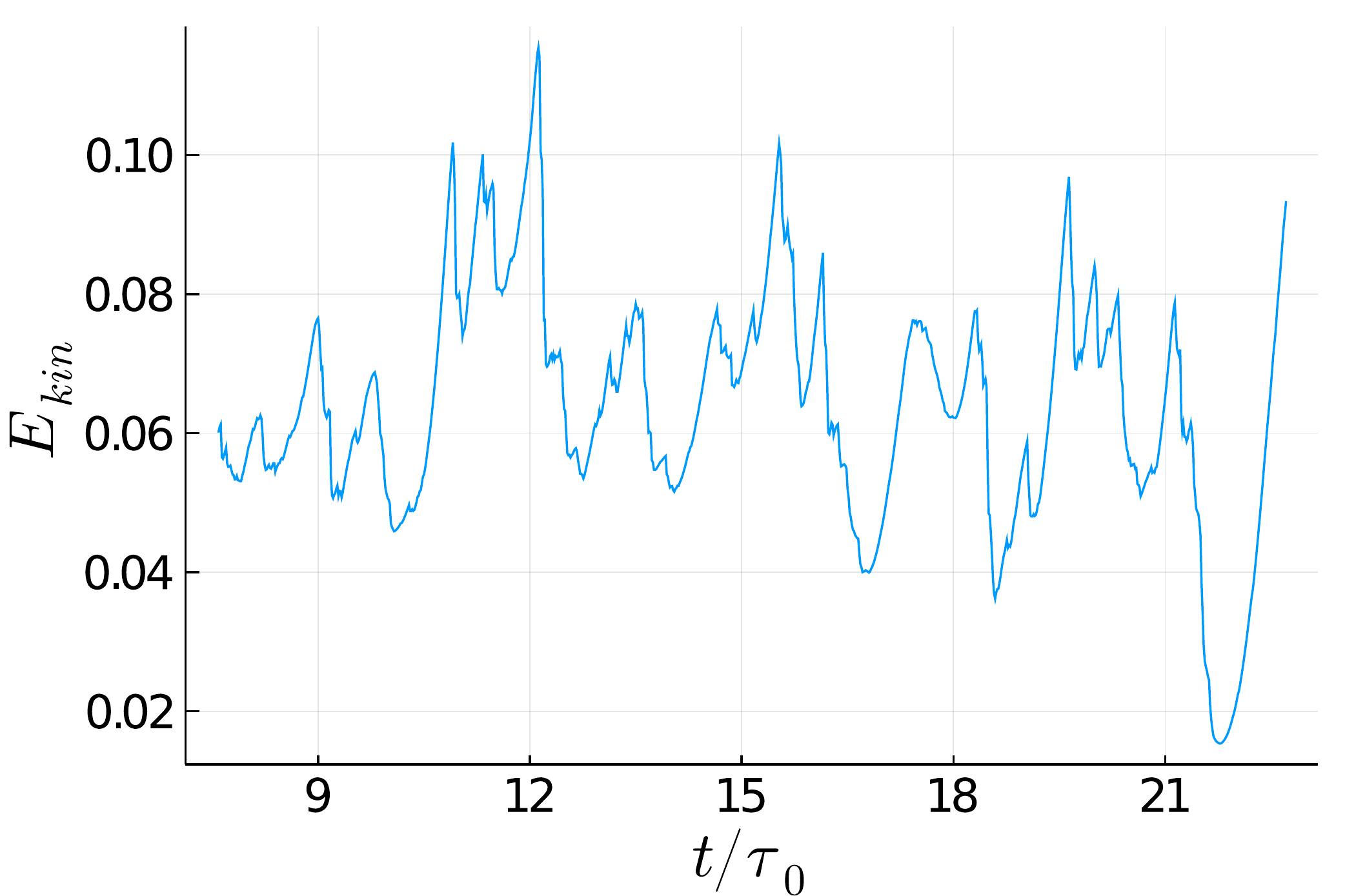}\hskip 0.5cm
\includegraphics[height=0.30\linewidth]{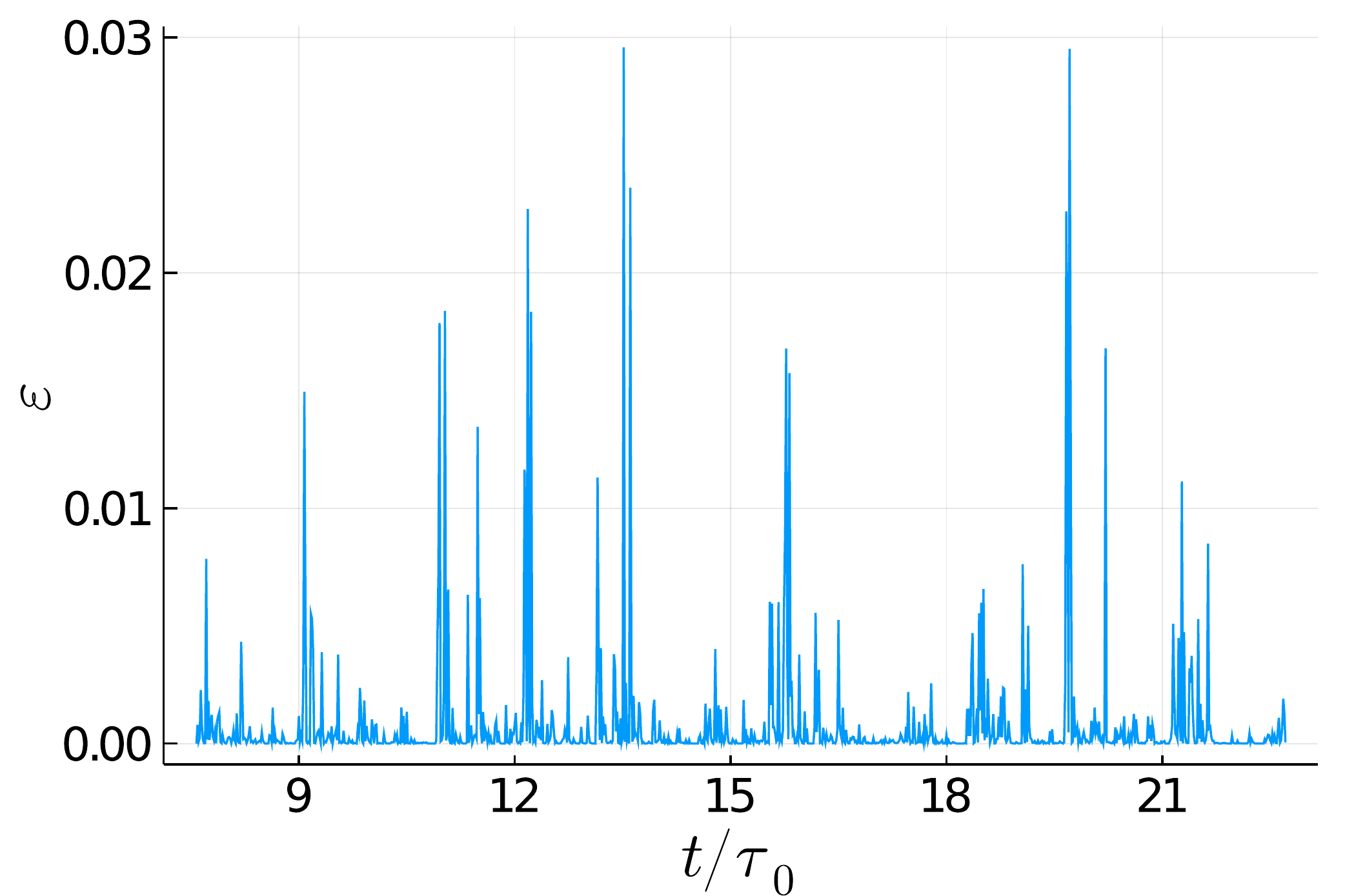}
\hskip 0.5cm
\includegraphics[height=0.30\linewidth]{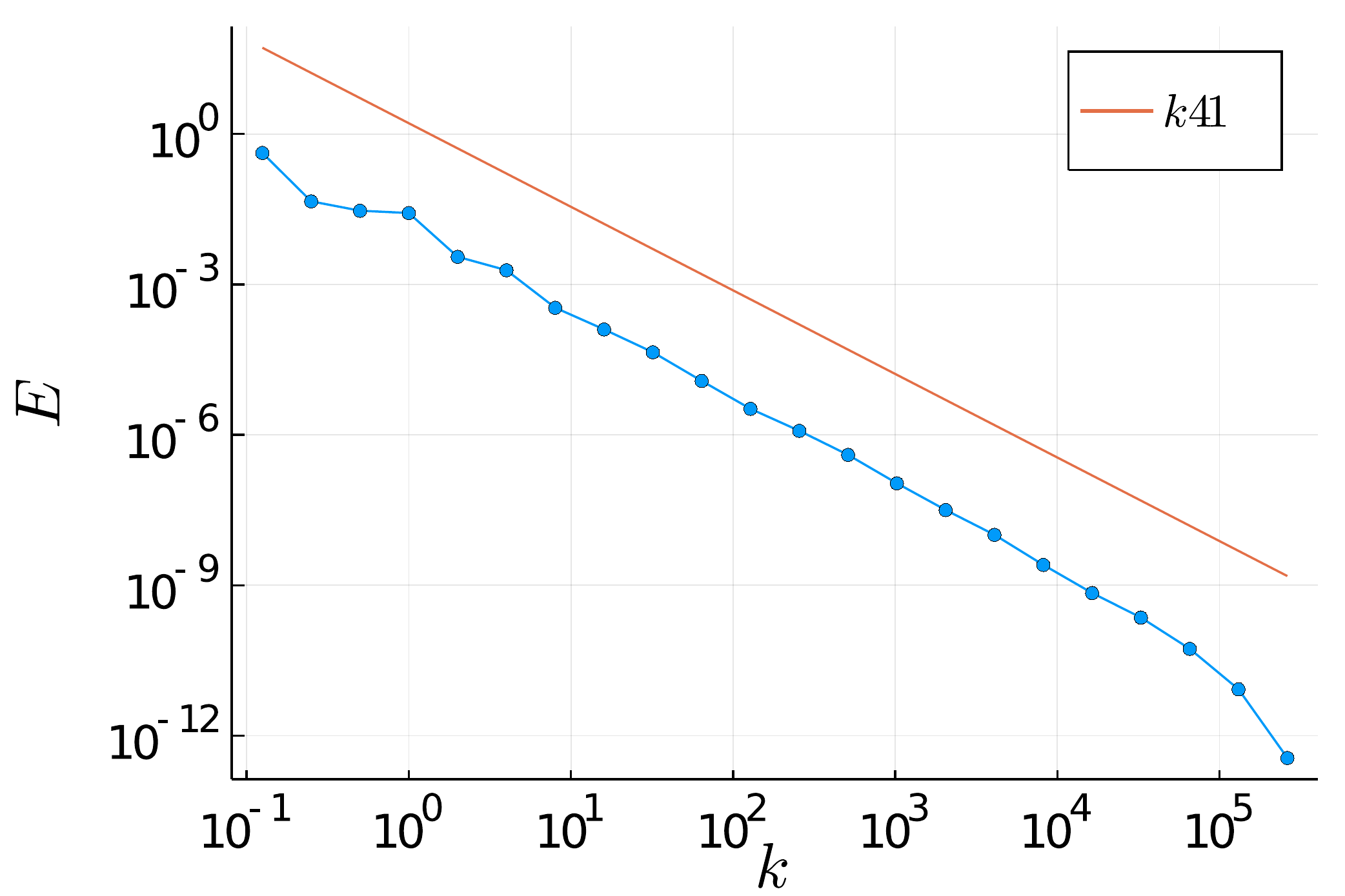}
\hskip 0.5cm
\includegraphics[height=0.30\linewidth]{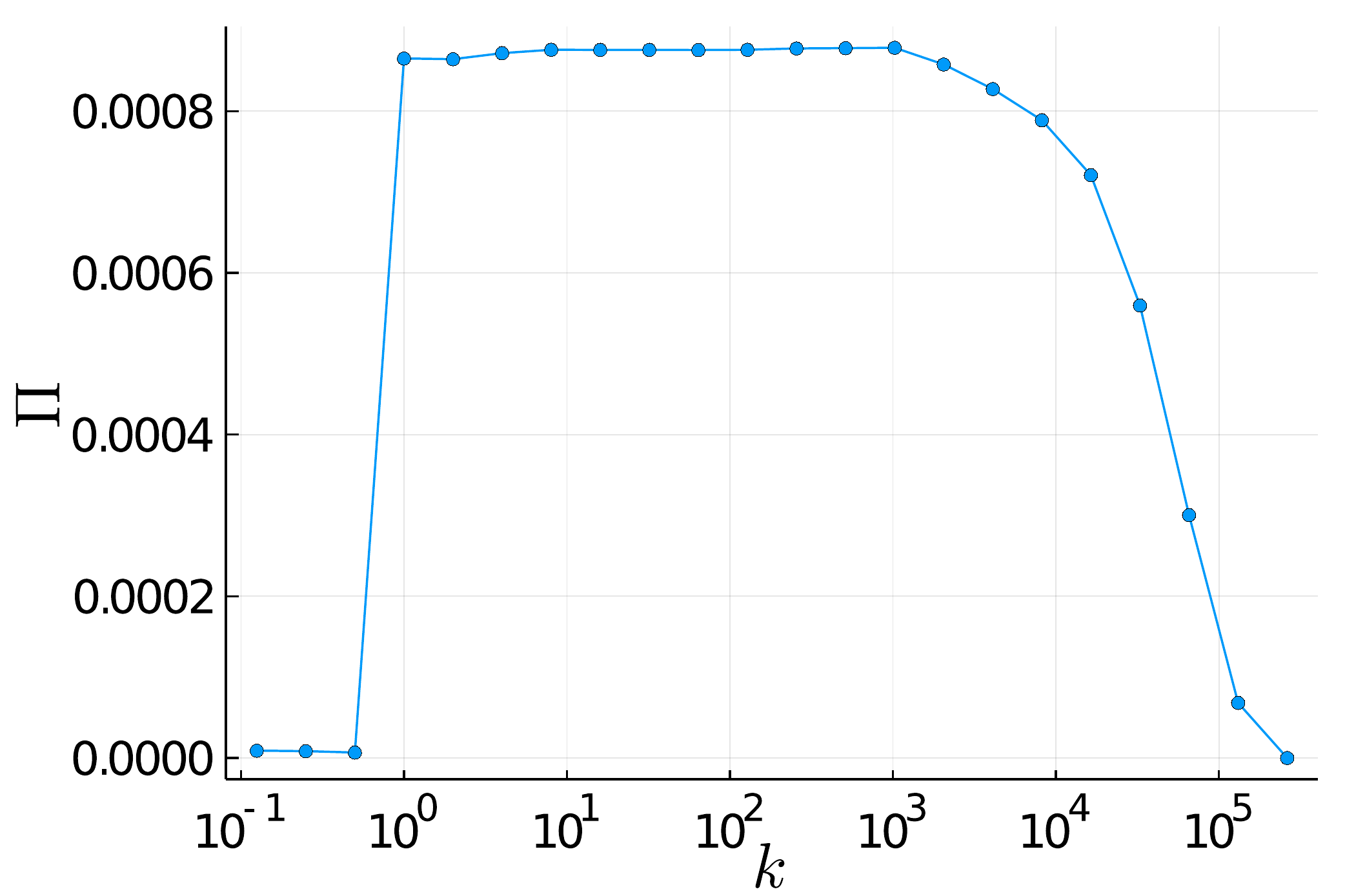}
\caption{Simulation results when learning $\theta$ in $M_i = \theta k_i^2 u_i$. Clockwise starting with the left upper graph: total kinetic energy, dissipation, energy flux, energy cascade}
\label{fig:GOY_ml_lr=1e-9}
\end{figure}

%We define our model as 
%\begin{equation}
%M_i=\theta k_i^2 u_i
%\label{eq:ml_model}
%\end{equation}
%with learnable parameter $\theta$. 
We substitute our diffusive model \eqref{eq:ml_model} into the differential equation~(\ref{eq:GOY_modified}), which we integrate
in the forward pass for $0.1t$. We then use the estimated velocity values $u_i$ to calculate the energy density given by equation~(\ref{eq:energy_density}) and subsequently the loss given by equation~(\ref{eq:objective}). As the energy density and consequently the loss is a statistical quantity we average over the last 1000 velocity vectors before calculating the derivative $\frac{\d L}{\d \theta}$ and updating the parameter $\theta$.
So we are attempting to learn a parameter for the viscosity $\nu$ so that the energy spectrum has the desired slope in the scaling range.
We start from a simulation with standard parameters at $t=500$ and set the initial value for $\theta$ to 0.0. Repeating this process for $\frac{1000}{\tau_0}$ normalized time steps leads to recovery of the desired slope of the energy cascade as well as the energy flux. This means that the learning algorithm learns on a window of $\frac{100}{\tau_0}$ normalized time steps into the immediate past of the system. The data window co-moves with the current simulation time and allows the learning algorithm to react and adapt to adverse back-reactions of the system to parameter changes. It is physically reasonable to set the window size to the characteristic time of the largest fluctuations which yields an estimate
of the period of time required to communicate changes of small-scale dissipation throughout the whole spectral range. The average large-eddy turn over time for a simulation with $\nu=10^{-8}$ is given by $\tau_0\approx 66.0 $. Our time window, including 1000 velocity vectors spaced at 0.1 time steps, thus covers $100t  \approx 1.5 \tau_0$. Simulation results are shown in Figure~\ref{fig:GOY_ml_lr=1e-9}. 

\begin{figure}
\centering
\includegraphics[height=0.3\linewidth]{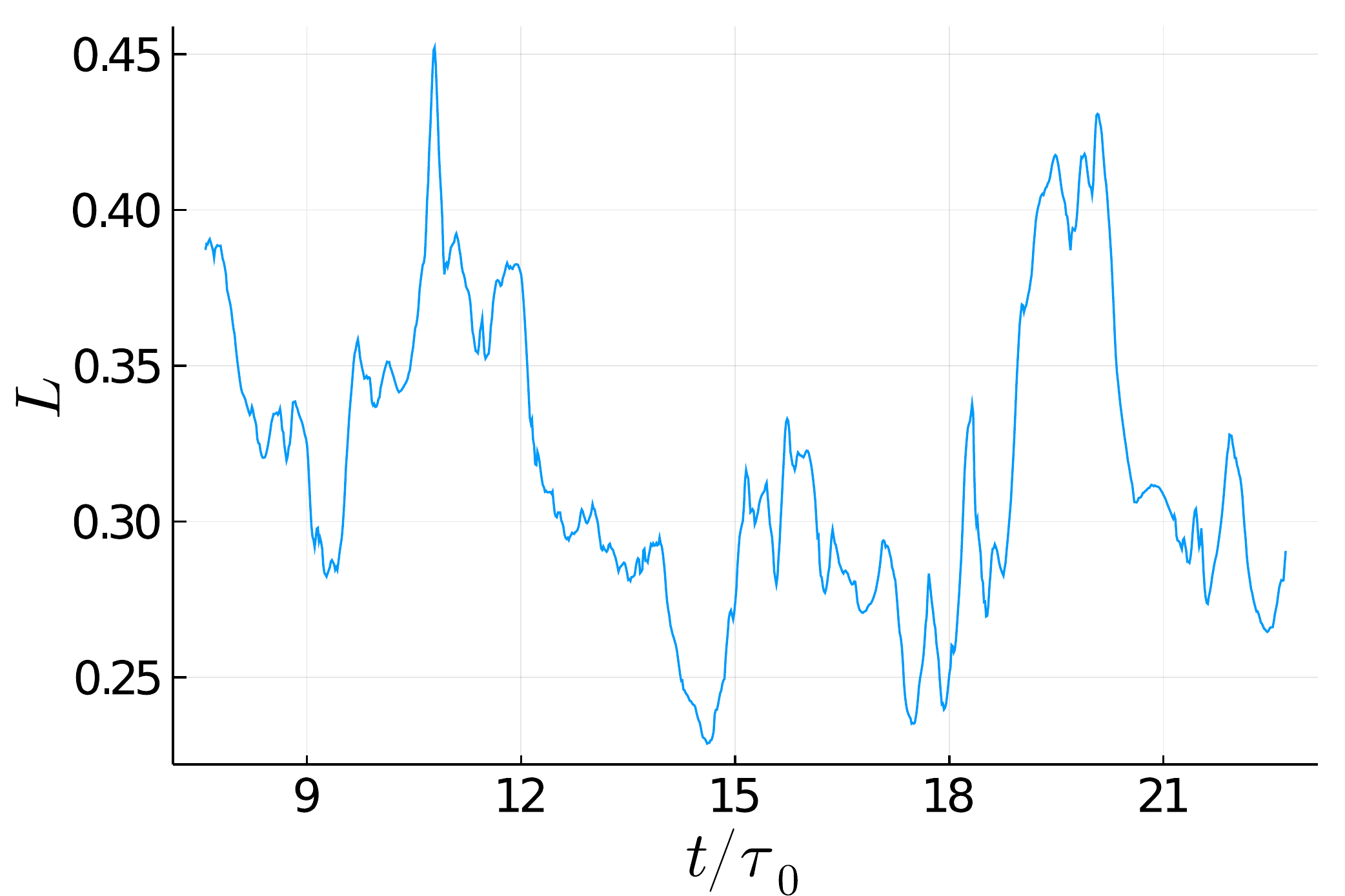}\hskip 0.5cm
\includegraphics[height=0.3\linewidth]{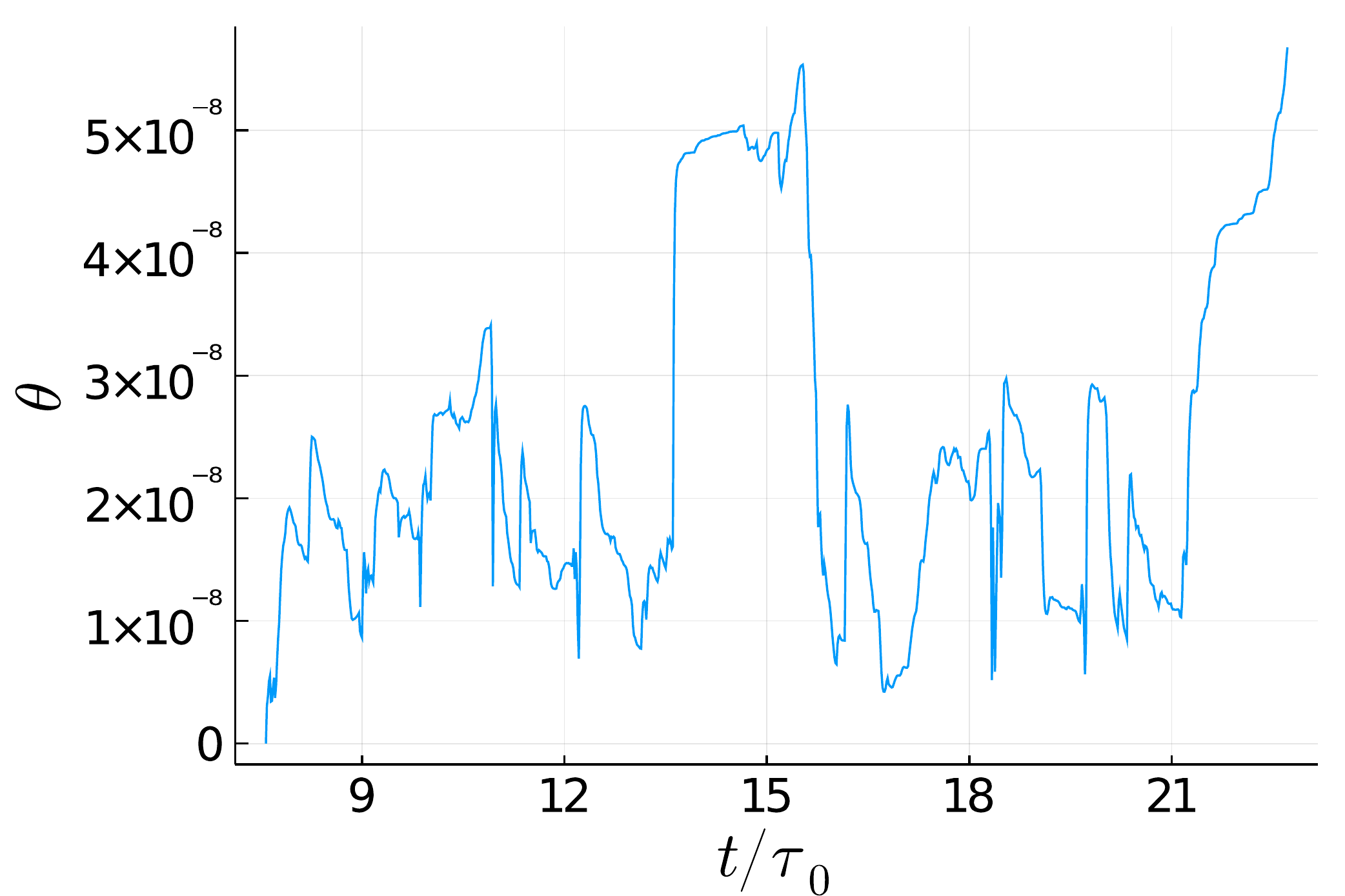}
\caption{Loss $L$ (left) and parameter $\theta$ (right) during optimization}
\label{fig:GOY_ml_lr=1e-9_loss_nu}
\end{figure}

In Figure~\ref{fig:GOY_ml_lr=1e-9_loss_nu} we show the evolution of our loss $L$ from Equation~(\ref{eq:objective}) and parameter $\theta$ during training. We do not start from a high loss since we continue a simulation with standard parameters for which the energy cascade already has the desired slope, so initially many of the 1000 velocity vectors over which we average stem from a simulation with a reasonable viscosity. However after 1000 update steps (or 100 time steps) the velocity vectors come entirely from solving the modified equations with the ML model. The parameter $\theta$ oscillates within one order of magnitude, but manages to reproduce the characteristic quantities that we observed in Section~\ref{sec:statistical_quantities}.

\section{Failure modes}\label{sec:failure_modes}
The success of the optimization described in Section~\ref{sec:learning_nu} is highly susceptible to the chosen learning rate as the order of magnitude of parameter $\theta$ is essential to reproduce the desired statistical quantities. Looking at the differential equation~(\ref{eq:GOY_modified}) two difficulties become apparent when defining $-M_i=-\theta k_i^2 u_i$. 
The term is of the form $\dot{y}= -c y$, for which the solution is of exponential form. For $\theta<0$ the true solution can become unstable since it grows exponentially. This effect can be especially strong since the larger shells do have large $k_i$. If we choose the learning rate too high and do a step so that $\theta$ becomes negative the numerical solver might not be able to calculate $\boldsymbol{u}^{(t+1)}$. So the optimizer cannot correct the mistake. On the other hand if $\theta$ is positive but large ($\geq 1\times 10^{-6}$) the equation becomes increasingly stiff. To keep the accuracy high the numerical solver has to decrease the stepsize. Since it takes a few steps until the effect becomes noticeable in the velocities and finally the loss, the stiffness can increase and might cause the solver to fail because it reaches the maximum number of steps. Especially the stiffness problem is caused by the delayed signal of the increasing loss and is not commonly found when using pre-computed velocity vectors as ground truth. The machine learning method is thus physically constrained in two aspects: i) the minimum size of the data window used to establish the proper gradient of the loss function, i.e. the next adjustment of the parameter $\theta$ and, ii) the possible range of the value of $\theta$ which is 
bounded by stiffness (time-step) considerations from above, as for $\theta\geq10^{-6}$ the solver steps required for integrating over $0.1$ time steps can exceed the default threshold of $10^5$, and which is bounded from below by zero since negative values would thwart the physically-motivated function of the learned term as an adaptable energy sink. We show additional results when optimizing for longer time periods within the above mentioned bounds in~\ref{appndx:additional_optimizations}.

\section{Conclusion}
In this paper we consider a modified version of the closure problem on a simplified model for turbulence. We include a machine learning term in the differential equation, which influences the numerical integration. The desired statistical quantities of turbulence serve as an objective. To calculate gradients in order to update our parameters we use the adjoint method.
With our work we present a different approach towards closure modeling with machine learning methods: Instead of finding the closure term using high fidelity DNS for supervised learning we require our model to reliably reproduce statistical properties of turbulence when incorporated into the differential equation. This problem formulation uncovers some pitfalls, unusual for machine learning tasks and which require the additional inclusion of (simple) physical principles to constrain the ML algorithm.
We regard the resulting scheme for the modelling of subgrid-scale NS turbulence as a proof-of-concept for a flexible physics-constrained ML technique that allows a  generalization of
methods like the dynamic procedure~\cite{germano1991, germano1992dynproc} and variations of it as well as self-similar extrapolation methods such as the approach discussed in~\cite{biferale2019selfsimsgs}.
The main advantage of the present approach is the increased level of flexibility within unavoidable physical constraints. This additional freedom permits increased flexibility with regard to the adaptive model term, e.g. its functional form and 
support (stencil) and with regard to the flow's statistical or any other properties that could be deemed to be important for SGS turbulence modelling.

\section*{Acknowledgments}
AKD is supported by the Research Training Group ``Differential Equation- and Data-driven Models in Life Sciences and Fluid Dynamics (DAEDALUS)'' (GRK 2433). KRM was supported in part by the German Ministry for Education and Research (BMBF)
under grants 01IS14013A-E, 01GQ1115, 01GQ0850, 01IS18056A, 01IS18025A and 01IS18037A, 
by the Information \& Communications Technology Planning \& Evaluation (IITP) grant funded by the Korea government (No. 2017-0-001779, , Artificial Intelligence Graduate School Program, Korea University), and Grant Math+, EXC 2046/1, Project ID 390685689 both funded by the German Research Foundation (DFG). We gratefully acknowledge very helpful discussions with John Platt.  

Correspondence to KRM and WCM.

\section*{Declaration of Competing Interest}
The authors declare that they have no known competing financial interests or personal relationships that influence the work reported in this paper.

\pagebreak
\FloatBarrier
\biboptions{sort&compress}
\bibliography{refs}
\bibliographystyle{unsrt}

\pagebreak
\appendix

\begin{center}
\Large{Appendix}
\end{center}

\tableofcontents
\unhidefromtoc
\section{Adjoint state method}\label{appndx:adjoint}

The adjoint state method is a well known practice that has been derived numerous times in different forms~\cite{pontryagin1987mathematical, bradley2013pde, Plessix2006review}. We include the derivation, following~\cite{vaibhav2020adjoint}, here merely for completeness.

We start out by formulating our objective as a constraint optimization problem

\begin{equation}
\min_\theta L(u(t))\quad \text{subject to}\quad F(\dot{u}, u, t, \theta)=\dot{u}-f(u, t, \theta)=0
\end{equation}
where we aim to find the set of parameters that minimize our loss (in our case that is the slope of the energy density over the shells). The constraint is just a differential equation describing the evolution of the velocity $u$

\begin{align*}
\dot{u} &= f(u, t, \theta)
\end{align*}

After having calculated the forward pass

\begin{align*}
u(t_1) = u_0 + \int_{t_0}^{t_1} \dot{u} \d t
\end{align*}
 we are interested in the gradient of the loss with respect to our parameters $\frac{\d L(u(t_1)}{\d \theta}$.
We can use Lagrange multipliers to reformulate a new objective
\begin{equation}
\mathcal{L}(\dot{u}, u, t, \theta) = L(u(t_1))-\int_{t_0}^{t_1}\lambda(t)^T\underbrace{F(\dot{u}, u, t, \theta)}_{= 0} \d t .
\end{equation}

As by construction, the constraint $F(\dot{u}, u, t, \theta)=0$ is always satisfied, we are free to chose $\lambda(t)$ however we want and $\frac{\d\mathcal{L}}{\d \theta}\equiv\frac{ \d L}{\d \theta}$

\begin{figure}
\centering
\includegraphics[width=0.5\linewidth]{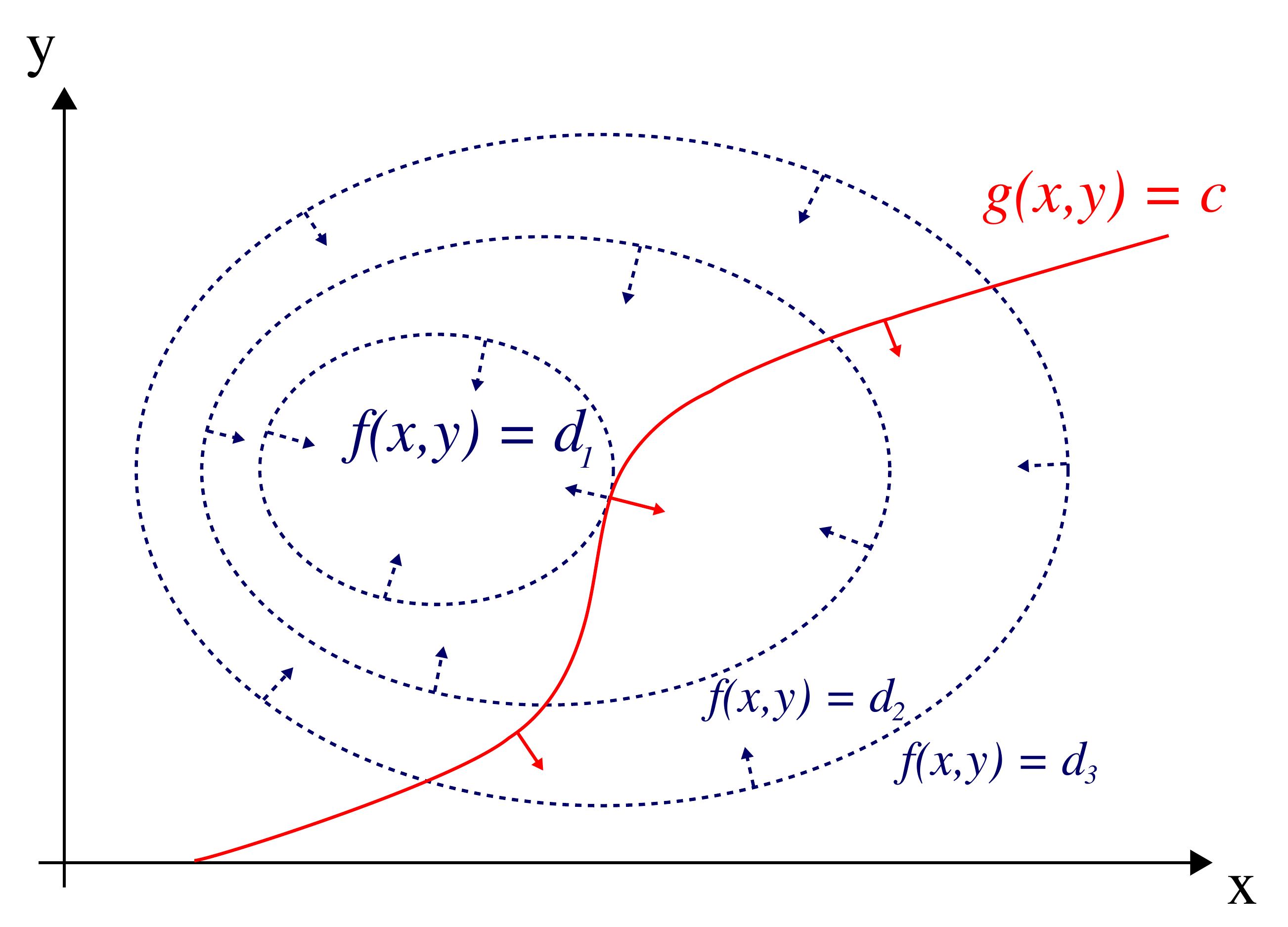}
\caption{The optimum for a constraint optimization problem is found at the point where the gradient of the function $f(x, y)$ that is optimized and the gradient of the constraint $g(x,y)=c$ lie on one line. At that point the constraint is tangential to the equipotential line of $f(x, y)$. Graphic from~\cite{wiki_lagrange} }
\label{appndx:fig:lagrange_multiplier}
\end{figure}

The idea behind Lagrange multipliers is that when we have a constraint optimization problem the minimum is the point where the gradient of the function that we are trying to minimize $L(u(t_1))$ and the gradient of the constraint $F(\dot{u}, u, t, \theta)=0$ align up to some multiplier, the Lagrange multiplier $\lambda$. See Figure~\ref{appndx:fig:lagrange_multiplier} for an intuition. This is exactly the equation we get when we take the derivative of the Lagrangian $\mathcal{L}$ with respect to our parameters $\theta$ and set it to 0.

\begin{equation}
\frac{\d\mathcal{L}}{\d \theta} = \frac{\d}{\d \theta}L(u(t_1))-\frac{\d}{\d \theta}\int_{t_0}^{t_1}\lambda(t)^T F(\dot{u}, u, t, \theta) \d t
\end{equation}

Now we focus on the integral part and use integration by parts

\begin{align*}
\int_{t_0}^{t_1}\lambda(t)^T F \d t &= \int_{t_0}^{t_1}\lambda(t)^T \left(\dot{u}-f(u, t, \theta)\right)\d t\\
&=\int_{t_0}^{t_1}\lambda(t)^T \dot{u} \d t - \int_{t_0}^{t_1}\lambda(t)^T f\d t\\
&=  \lambda(t)^T  u(t)\bigg\vert_{t_0}^{t_1} - \int_{t_0}^{t_1} \dot{\lambda}(t)^T  u(t) \d t - \int_{t_0}^{t_1}\lambda(t)^T f\d t\\
&=\lambda(t_1)^T  u(t_1)-\lambda(t_0)^T  u(t_0)- \int_{t_0}^{t_1} \dot{\lambda}(t)^T u(t) \d t - \int_{t_0}^{t_1}\lambda(t)^T f\d t\\
&= \lambda(t_1)^T  u(t_1)-\lambda(t_0)^T  u(t_0)-\int_{t_0}^{t_1}\left( \dot{\lambda}^T u+\lambda^T  f\right) \d t
\end{align*}
taking the derivative with respect to $\theta$ gives

\begin{align*}
\frac{\d}{\d \theta}\int_{t_0}^{t_1}\lambda(t)^T F \d t &= \lambda(t_1)^T \frac{\d u(t_1)}{\d\theta}-0-\int_{t_0}^{t_1}\left( \dot{\lambda}^T \frac{\d u}{\d\theta} +\lambda^T  \frac{\d f}{\d \theta}\right) \d t
\end{align*}

and applying the chain rule $\frac{\d f}{\d \theta} = \frac{\partial f}{\partial\theta}+ \frac{\partial f}{\partial z}\frac{\d u}{\d \theta}$ gives

\begin{align*}
\frac{\d}{\d \theta}\int_{t_0}^{t_1}\lambda(t)^T F \d t &= \lambda(t_1)^T  \frac{\d u(t_1)}{\d\theta}-0-\int_{t_0}^{t_1} \left(\dot{\lambda}^T +\lambda^T \frac{\partial f}{\partial u}\right)\frac{\d u}{\d \theta} -\int_{t_0}^{t_1}\frac{\partial f}{\partial \theta} \d t
\end{align*}

so the derivative of the Lagrangian is then

\begin{align*}
\frac{\d\mathcal{L}}{\d \theta} &= \frac{\d}{\d \theta}L(u(t_1))-\frac{\d}{\d \theta}\int_{t_0}^{t_1}\lambda(t)^T F(\dot{u}, u, t, \theta) \d t\\
&=\frac{\partial L}{\partial u(t_1)}\frac{\d u(t_1)}{\d\theta} - \frac{\d}{\d \theta}\int_{t_0}^{t_1}\lambda(t)^T F(\dot{u}, u, t, \theta) \d t\\
&=\left( \frac{\partial L}{\partial u(t_1)}-\lambda(t_1)^T \right)\frac{\d u(t_1)}{\d\theta}-\int_{t_0}^{t_1} \left(\dot{\lambda}^T +\lambda^T \frac{\partial f}{\partial u}\right)\frac{\d u}{\d \theta} -\int_{t_0}^{t_1}\frac{\partial f}{\partial \theta} \d t
\end{align*}

In this equation some derivatives, like $\frac{\d u}{\d \theta}$ are hard to calculate. So we choose $\lambda(t)$ in a way that they drop out. 
\begin{equation}
\lambda(t) = \frac{\partial L}{\partial u} 
\end{equation}
 is then called the adjoint state with derivative 
 \begin{equation}
 \dot{\lambda} = -\lambda\frac{\partial f}{\partial u}
 \end{equation}
 
We can get the derivative $\dot{\lambda}$ from the definition with finite differences when using the chain rule $\lambda(t) = \frac{\partial L}{\partial u(t)}  =\frac{\partial L}{\partial u(t+\epsilon)} \frac{\partial u(t+\epsilon)}{\partial u(t)}  = \lambda(t+\epsilon) \frac{\partial u(t+\epsilon)}{\partial u(t)}$

\begin{align*}
\dot{\lambda} &= \lim_{\epsilon \to 0} \frac{\lambda(t+\epsilon)-\lambda(t)}{\epsilon}\\
&=\lim_{\epsilon \to 0} \frac{\lambda(t+\epsilon)-\lambda(t+\epsilon) \frac{\partial u(t+\epsilon)}{\partial u(t)}}{\epsilon}\qquad \qquad\bigg\rvert\text{Taylor around $u(t)$}\\
&=\lim_{\epsilon \to 0}\frac{1}{\epsilon} \left(\lambda(t+\epsilon)-\lambda(t+\epsilon)\frac{\partial}{\partial u}\left[ u+ \epsilon\dot{u} + O(\epsilon^2)\right]\right)\\
&=\lim_{\epsilon \to 0}\frac{1}{\epsilon} \left(\lambda(t+\epsilon)-\lambda(t+\epsilon)\left[ I+ \epsilon\frac{\partial f}{\partial u} + O(\epsilon^2)\right]\right)\\
&=\lim_{\epsilon \to 0}\frac{1}{\epsilon} \left(\lambda(t+\epsilon)\epsilon\frac{\partial f}{\partial u} + O(\epsilon^2)\right)\\
&=\lim_{\epsilon \to 0}\left(\lambda(t+\epsilon)\frac{\partial f}{\partial u} + O(\epsilon)\right)\\
&=\lambda(t)\frac{\partial f}{\partial u}
\end{align*}

 Using this, the derivative of the Lagrangian simplifies to 
\begin{equation}
\frac{\d\mathcal{L}}{\d \theta} =\frac{\d L}{\d \theta}= -\int_{t_1}^{t_0} \lambda(t)^T \frac{\partial f}{\partial \theta}
\end{equation}

This integral can be solved backwards in time using a numerical solver. So the derivative of the loss function with respect to the parameters can be calculated with a desired precision and then used by an optimizer to change the parameters $\theta$.

\FloatBarrier

\section{Additional optimizations}\label{appndx:additional_optimizations}
\begin{figure}
\centering
\includegraphics[height=0.3\linewidth]{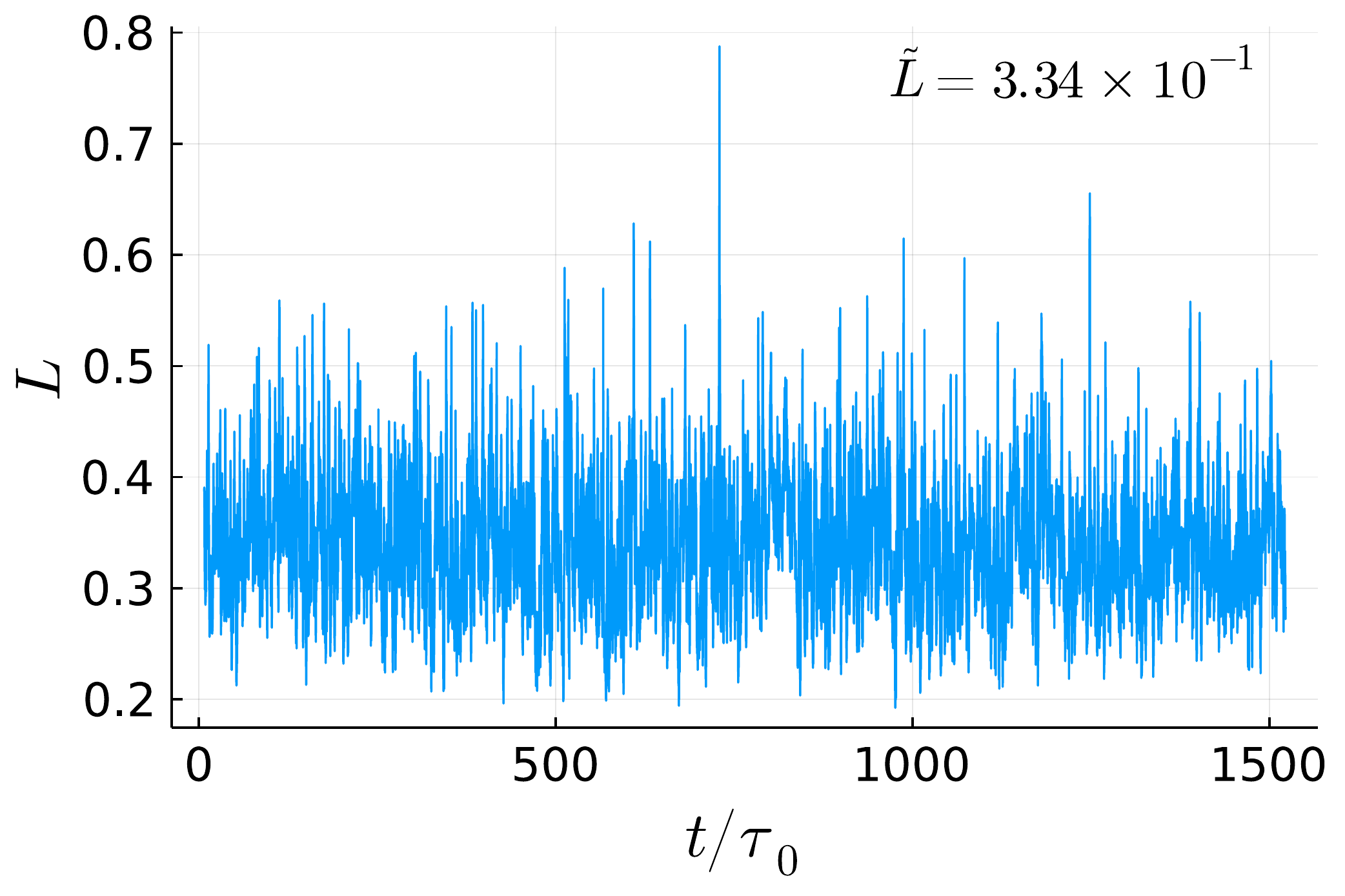}\hskip 0.5cm
\includegraphics[height=0.3\linewidth]{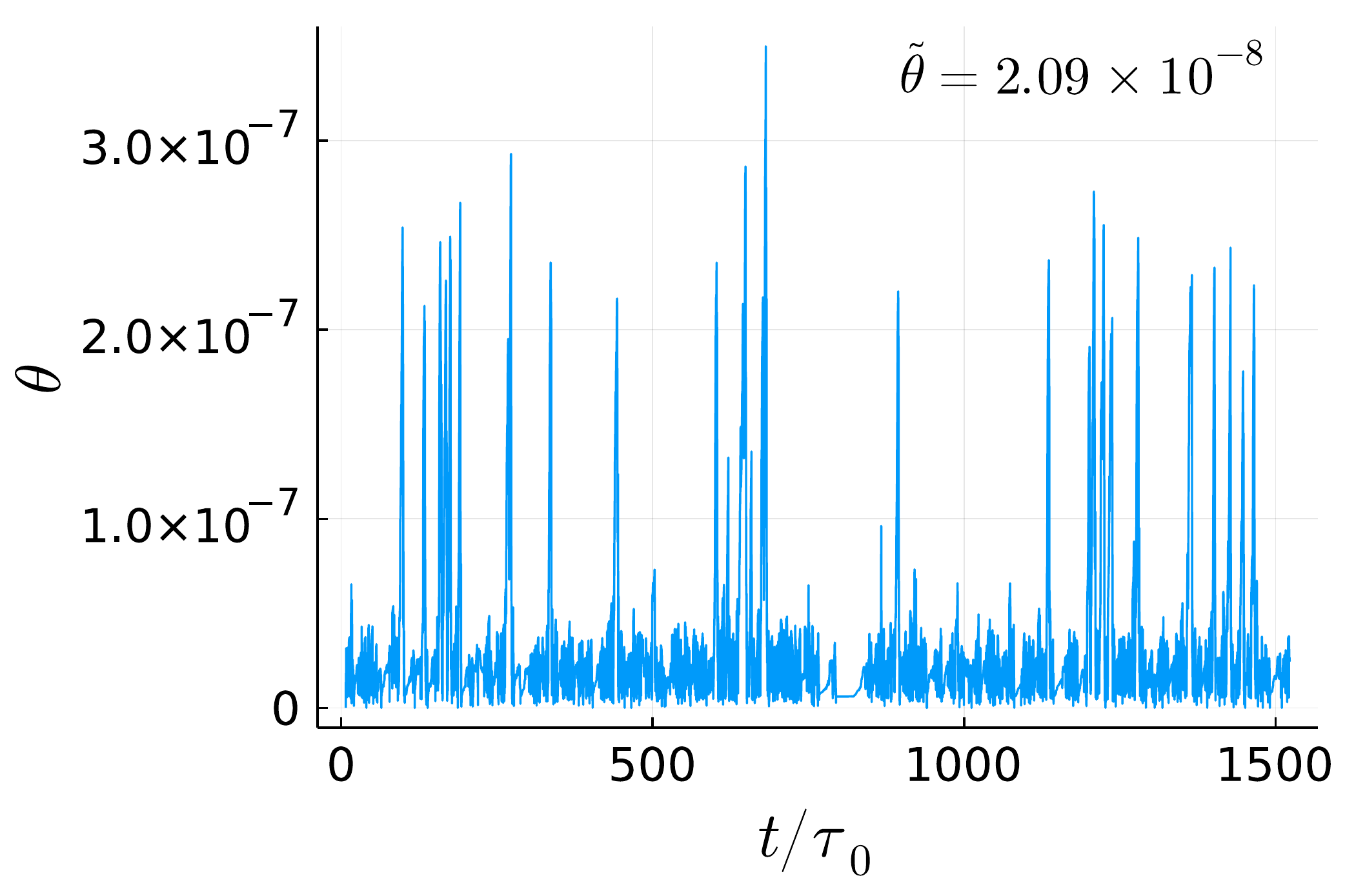}
\caption{Loss $L$ (left) and parameter $\theta$ (right) during optimization with learning rate $1\times10^{-9}$ when we update parameter~$\theta$ every 0.1 time steps. Median values are depicted as $\tilde{L}$ and $\tilde{\theta}$ respectively.}
\label{fig:GOY_ml_lr=1e-9_update_freq=1}
\end{figure}

\begin{figure}
\centering
\includegraphics[height=0.3\linewidth]{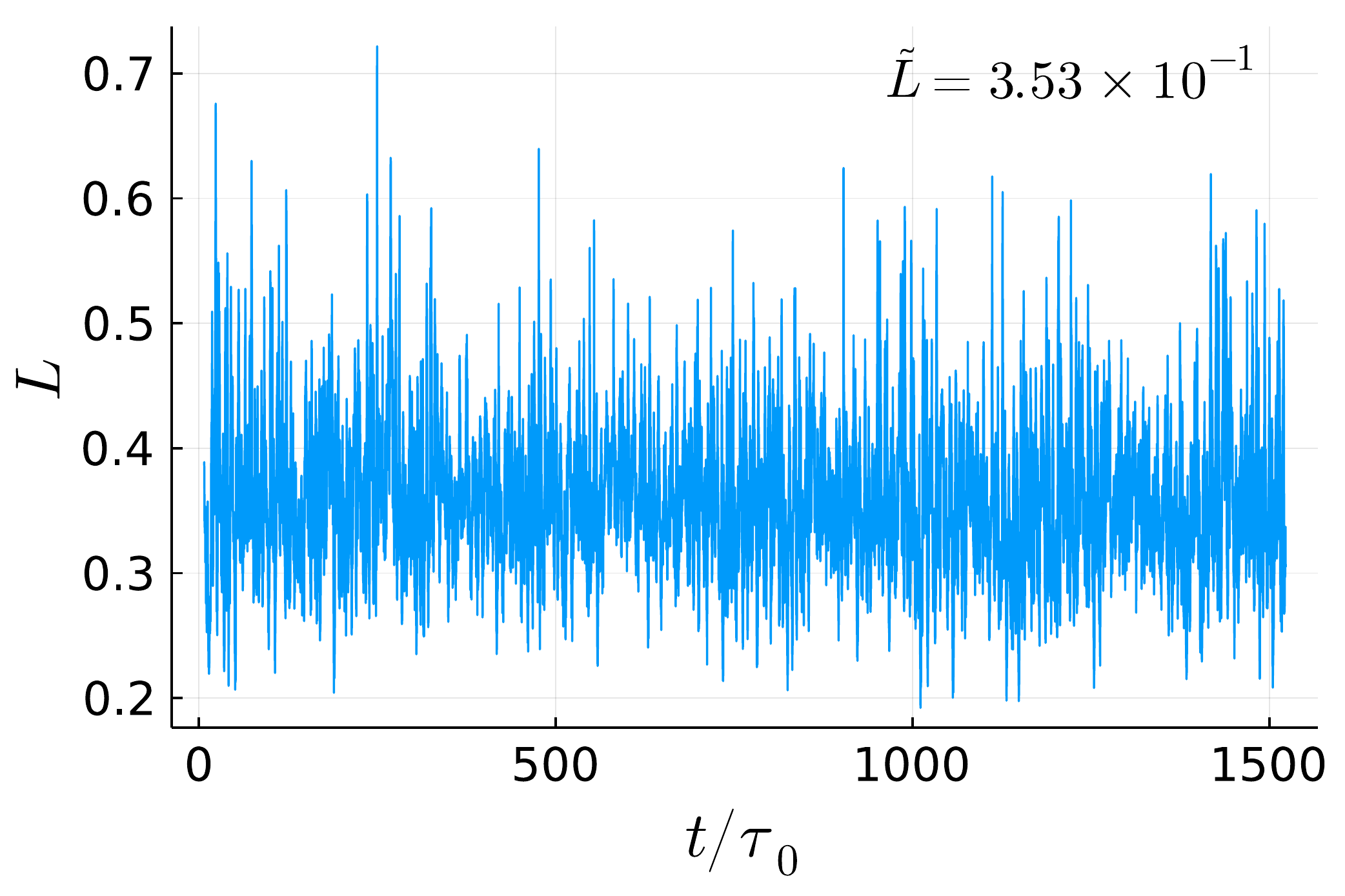}\hskip 0.5cm
\includegraphics[height=0.3\linewidth]{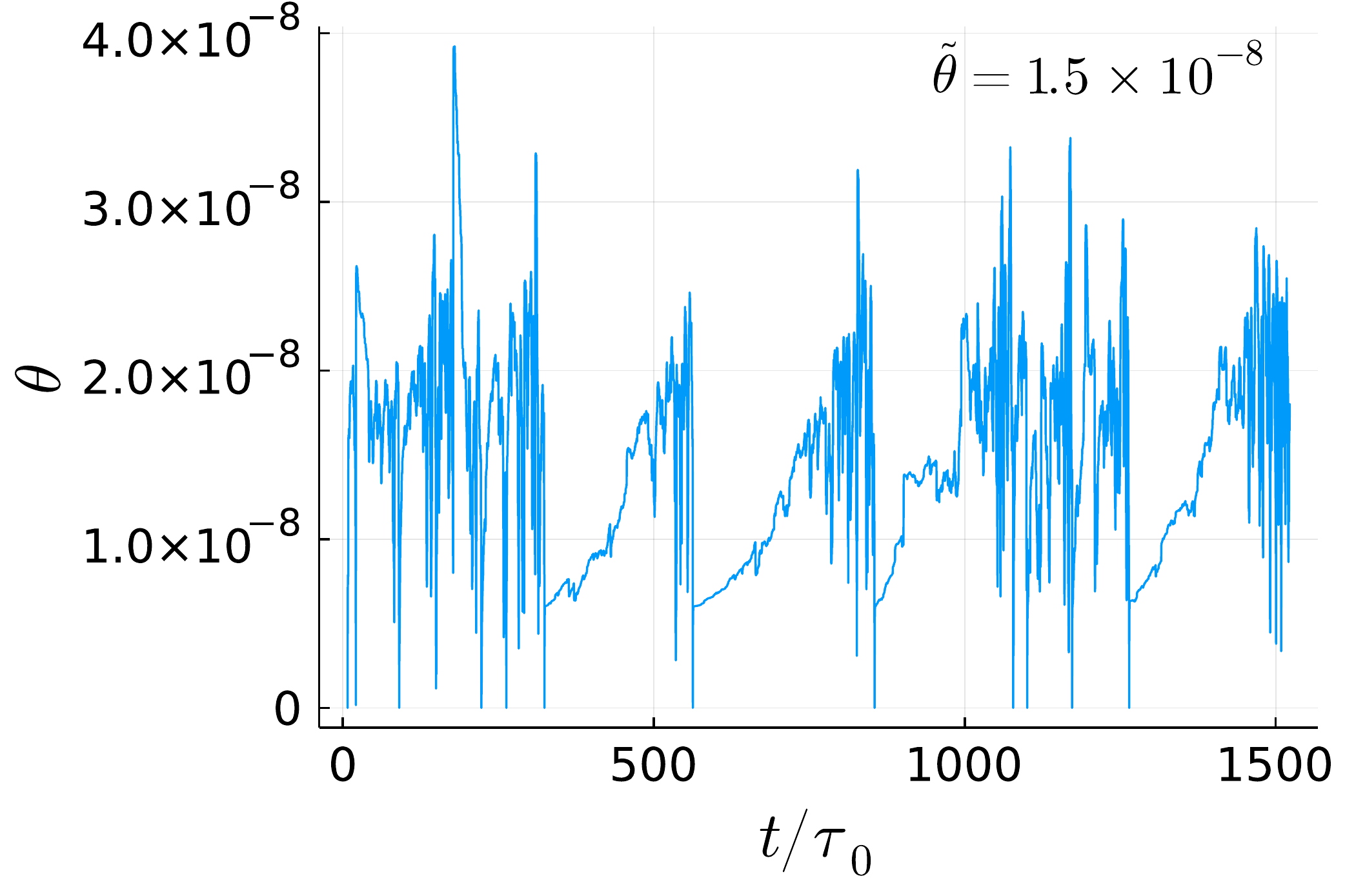}
\caption{Loss $L$ (left) and parameter $\theta$ (right) during optimization with learning rate $1\times10^{-9}$ when we update parameter~$\theta$ every time step. Median values are depicted as $\tilde{L}$ and $\tilde{\theta}$ respectively.}
\label{fig:GOY_ml_lr=1e-9_update_freq=10}
\end{figure}

\begin{figure}
\centering
\includegraphics[height=0.3\linewidth]{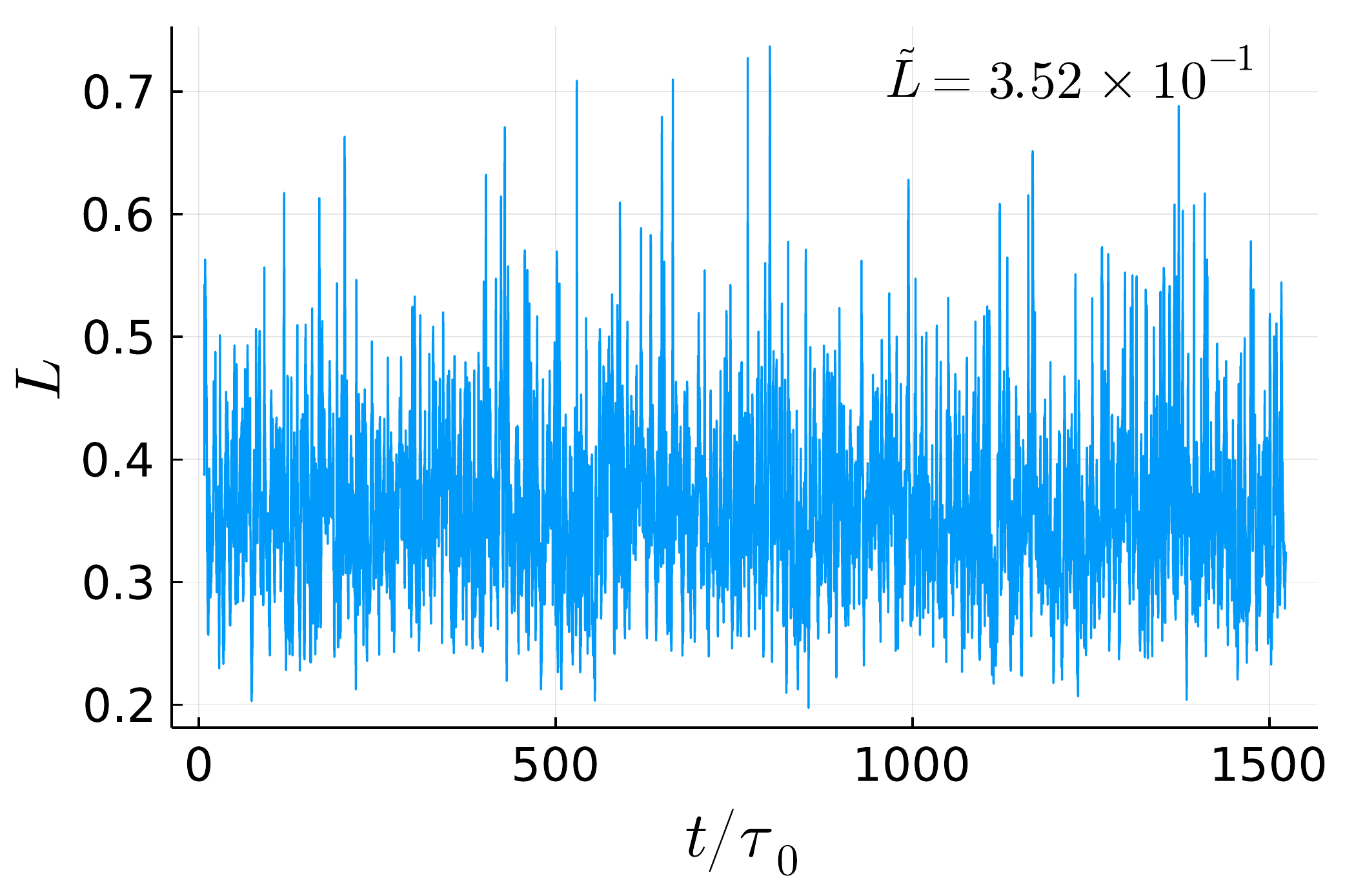}\hskip 0.5cm
\includegraphics[height=0.3\linewidth]{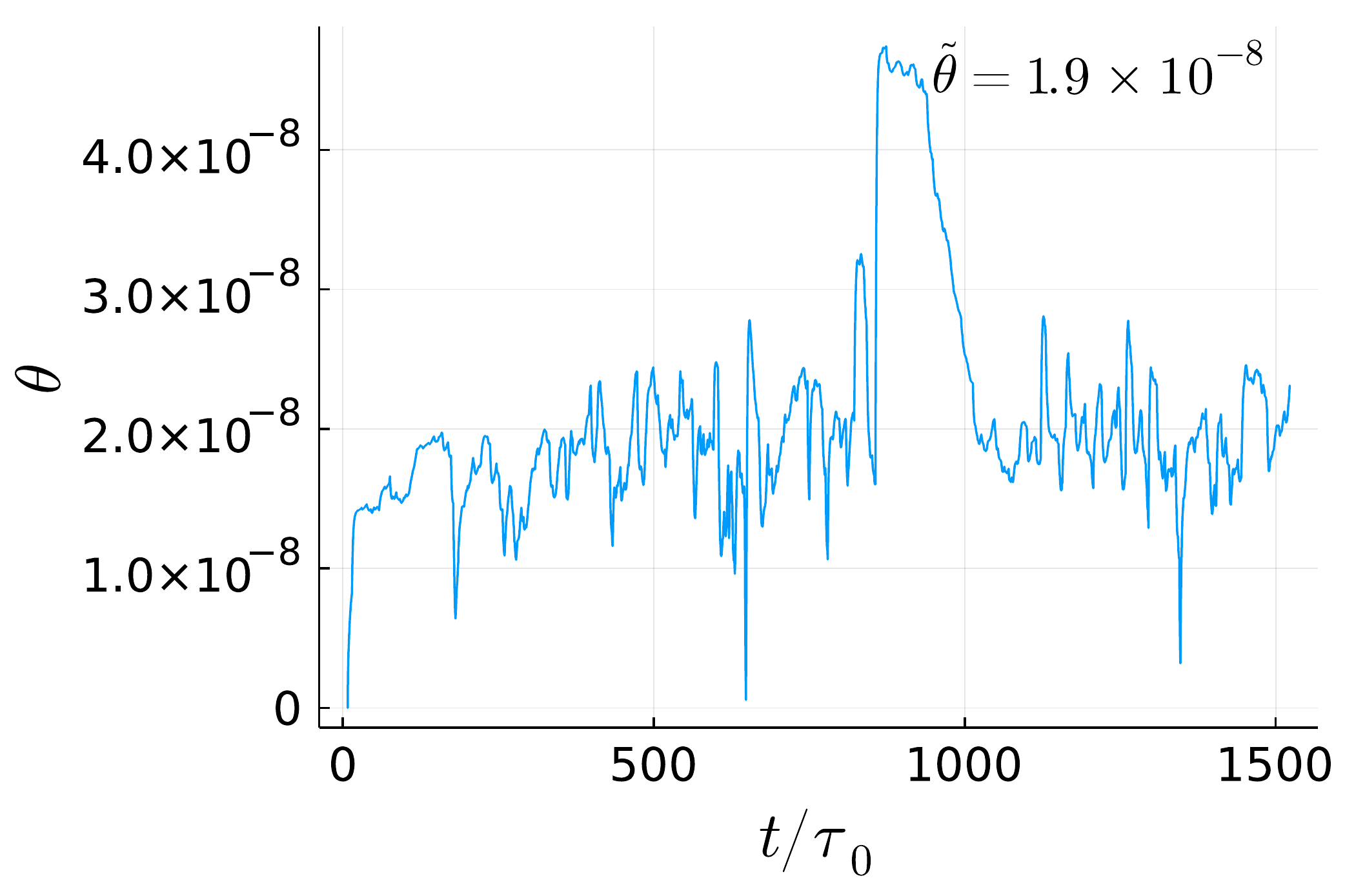}
\caption{Loss $L$ (left) and parameter $\theta$ (right) during optimization with learning rate $1\times10^{-9}$ when we update parameter~$\theta$ every 10 time steps. Median values are depicted as $\tilde{L}$ and $\tilde{\theta}$ respectively.}
\label{fig:GOY_ml_lr=1e-9_update_freq=100}
\end{figure}

\begin{figure}
\centering
\includegraphics[height=0.3\linewidth]{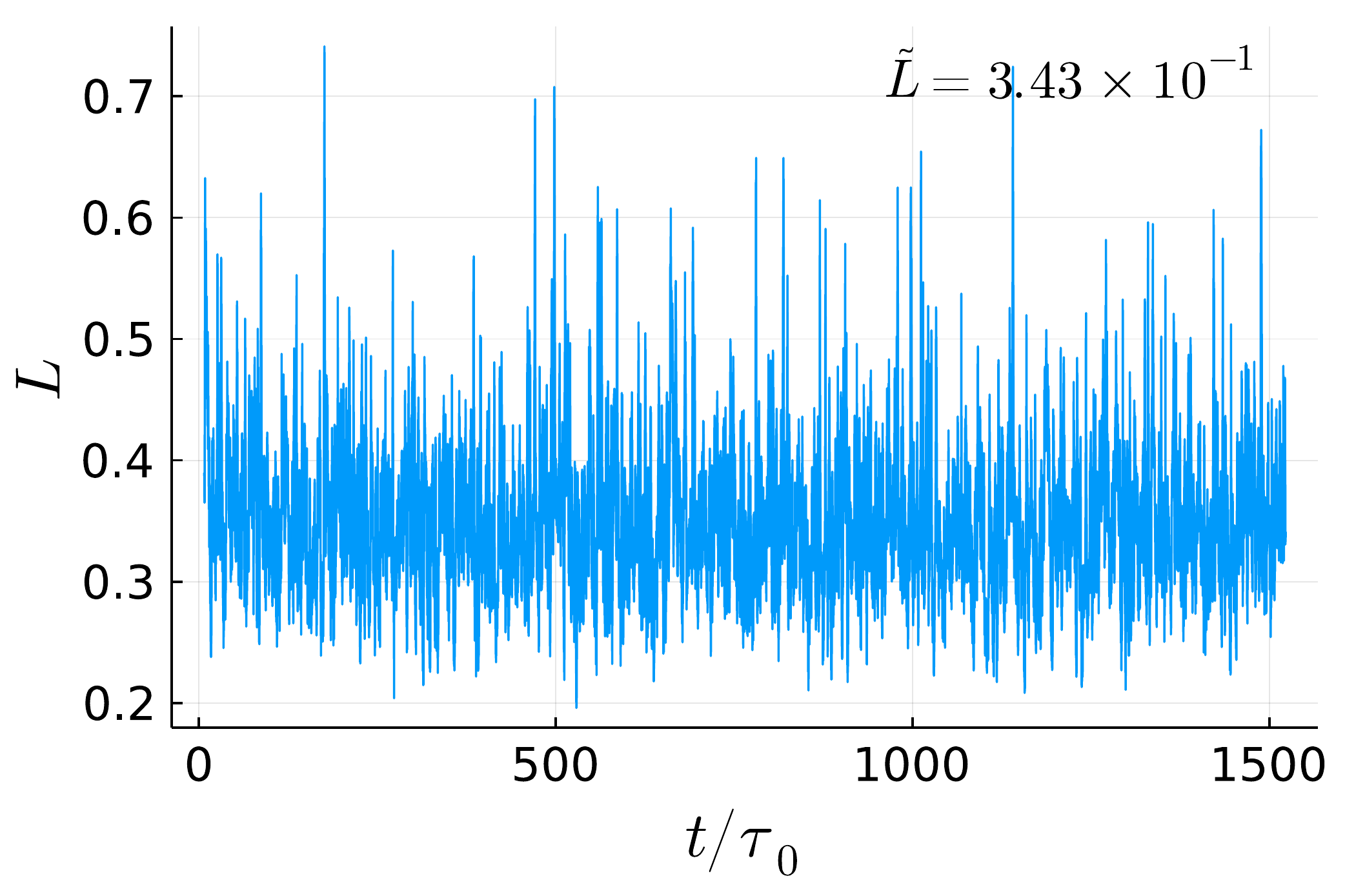}\hskip 0.5cm
\includegraphics[height=0.3\linewidth]{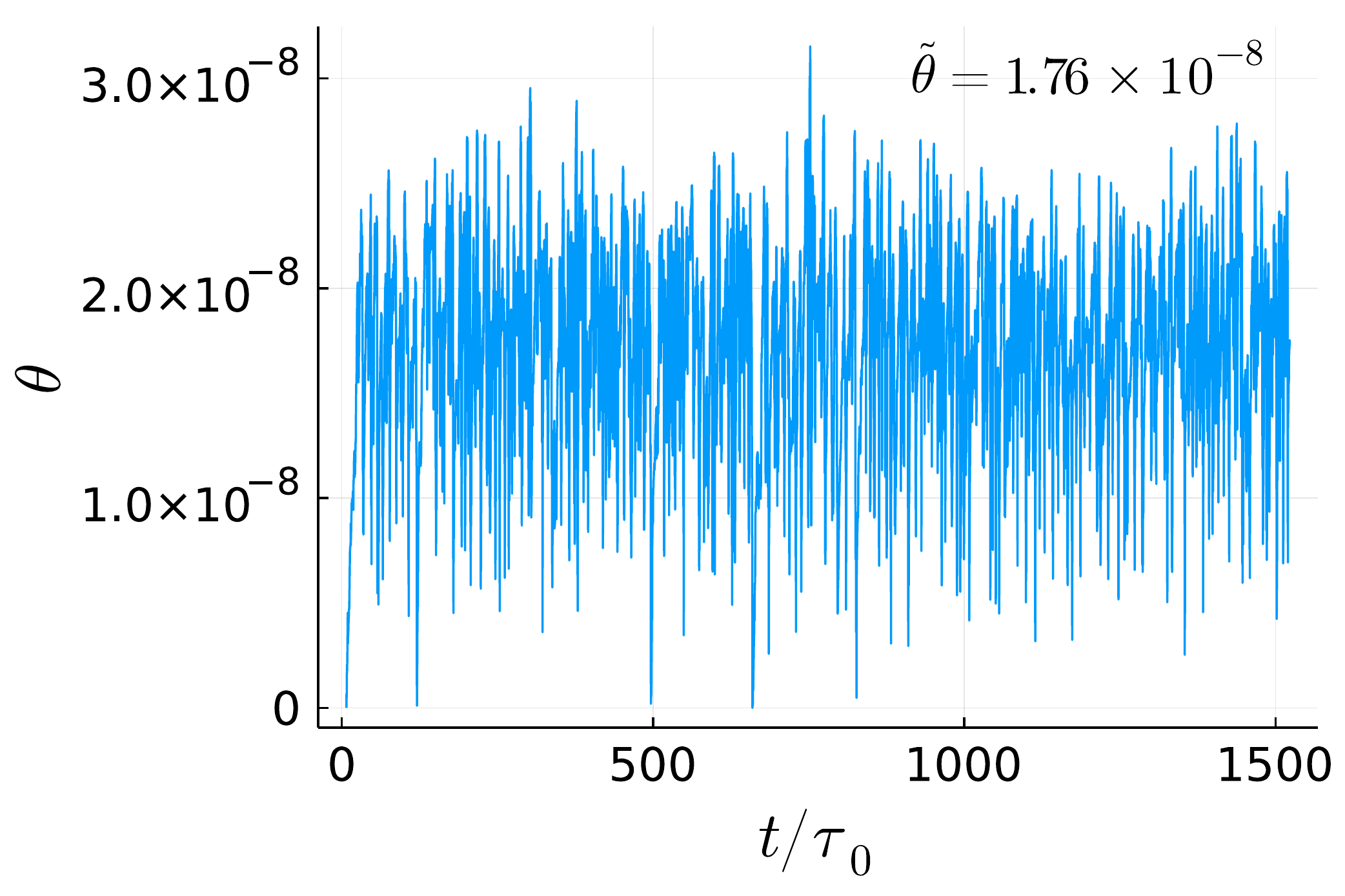}
\caption{Loss $L$ (left) and parameter $\theta$ (right) during optimization with learning rate $1\times10^{-10}$ when we update parameter~$\theta$ every 0.1 time steps. Median values are depicted as $\tilde{L}$ and $\tilde{\theta}$ respectively.}
\label{fig:GOY_ml_lr=1e-10_update_freq=1}
\end{figure}

\begin{figure}
\centering
\includegraphics[height=0.3\linewidth]{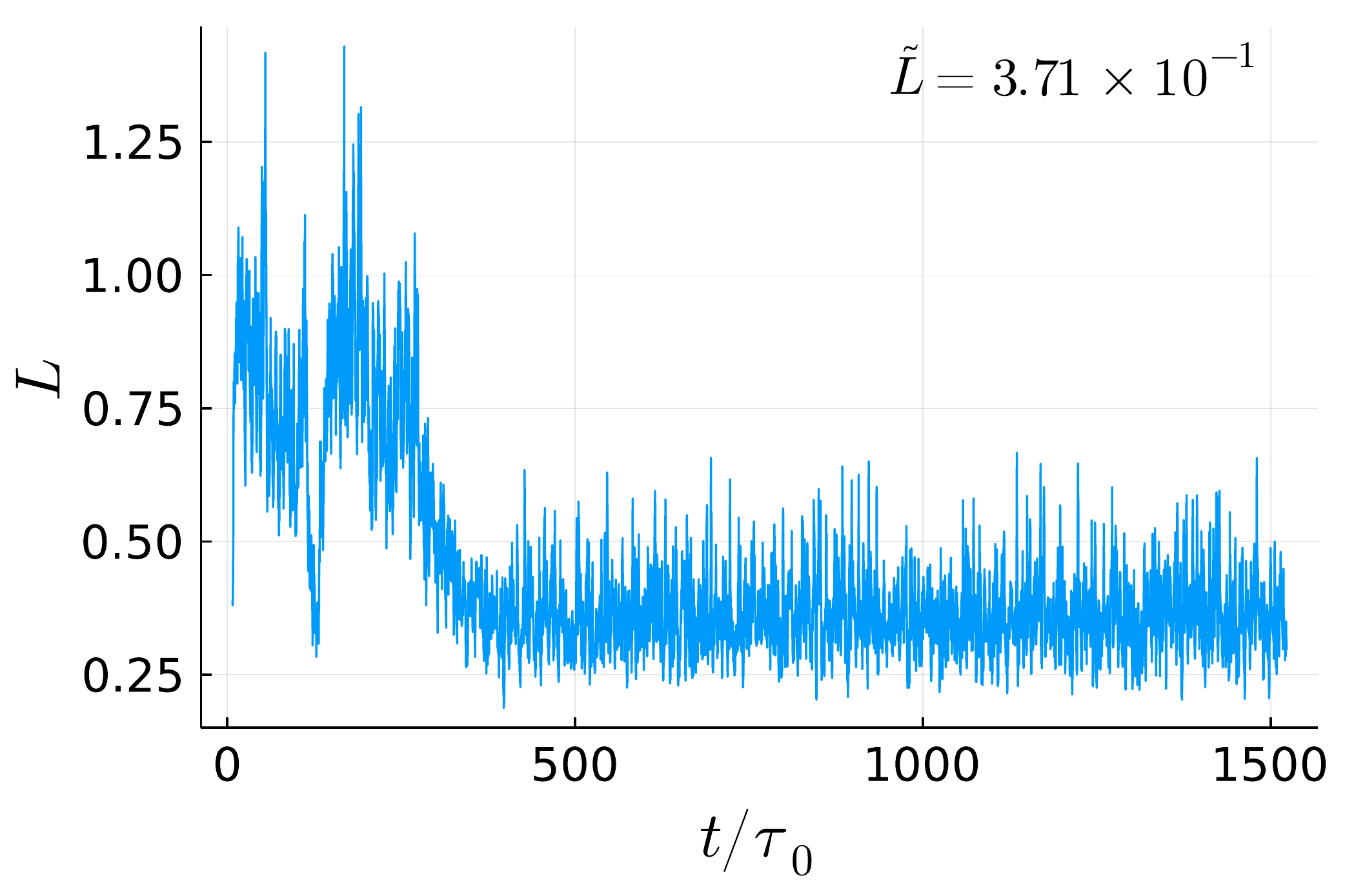}\hskip 0.5cm
\includegraphics[height=0.3\linewidth]{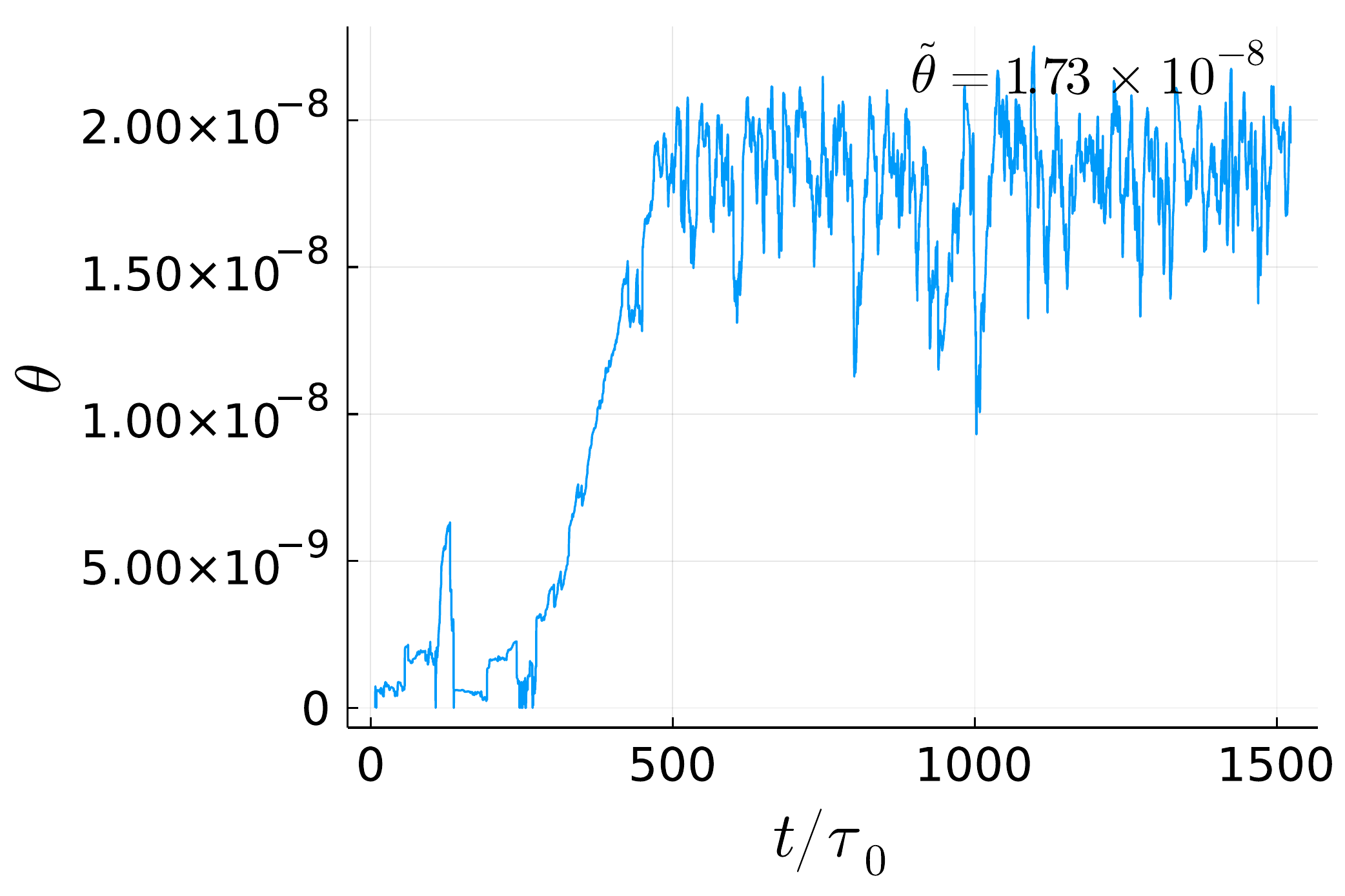}
\caption{Loss $L$ (left) and parameter $\theta$ (right) during optimization with learning rate $1\times10^{-10}$ when we update parameter~$\theta$ every time step. Median values are depicted as $\tilde{L}$ and $\tilde{\theta}$ respectively.}
\label{fig:GOY_ml_lr=1e-10_update_freq=10}
\end{figure}

\begin{figure}
\centering
\includegraphics[height=0.3\linewidth]{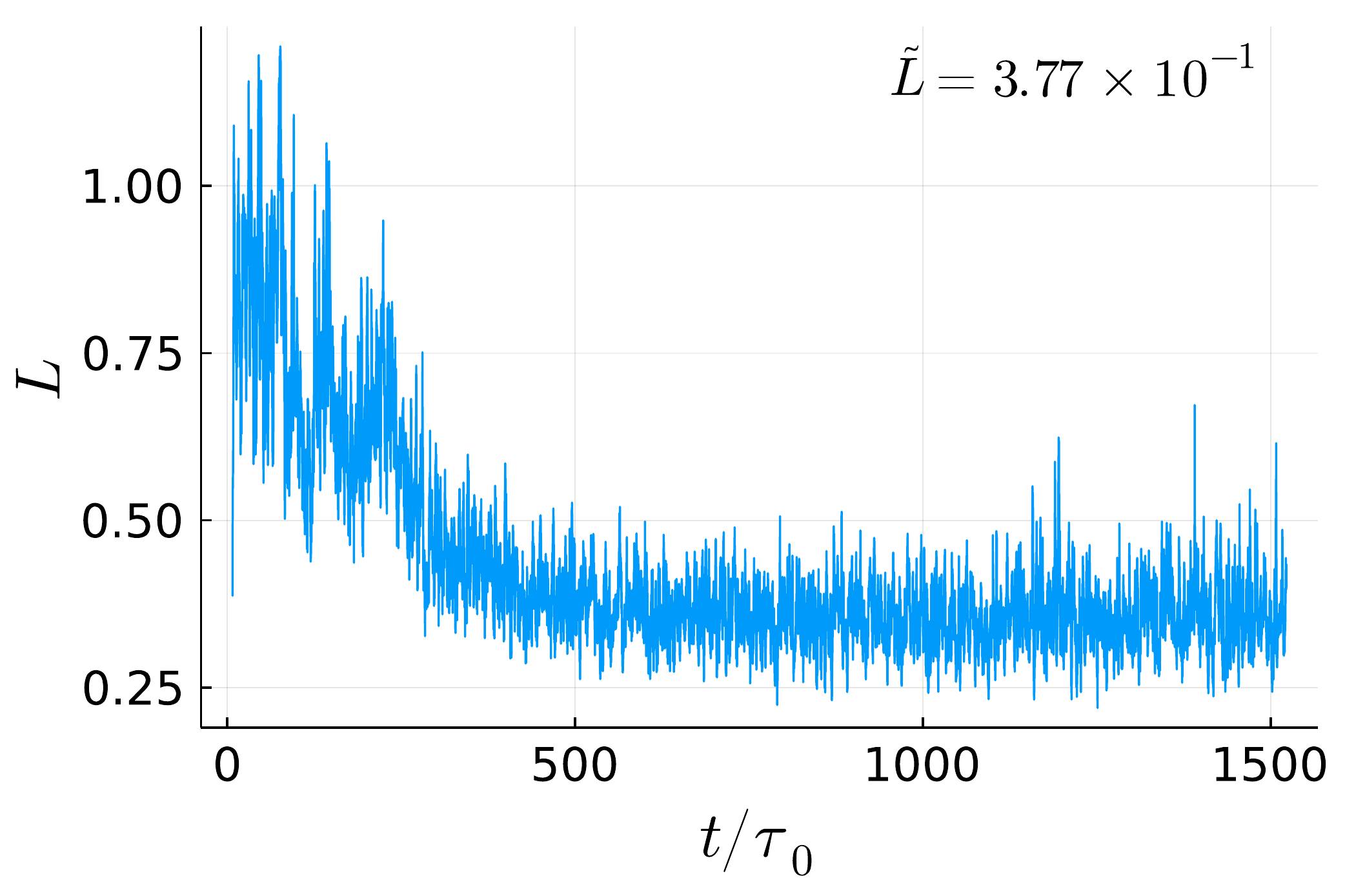}\hskip 0.5cm
\includegraphics[height=0.3\linewidth]{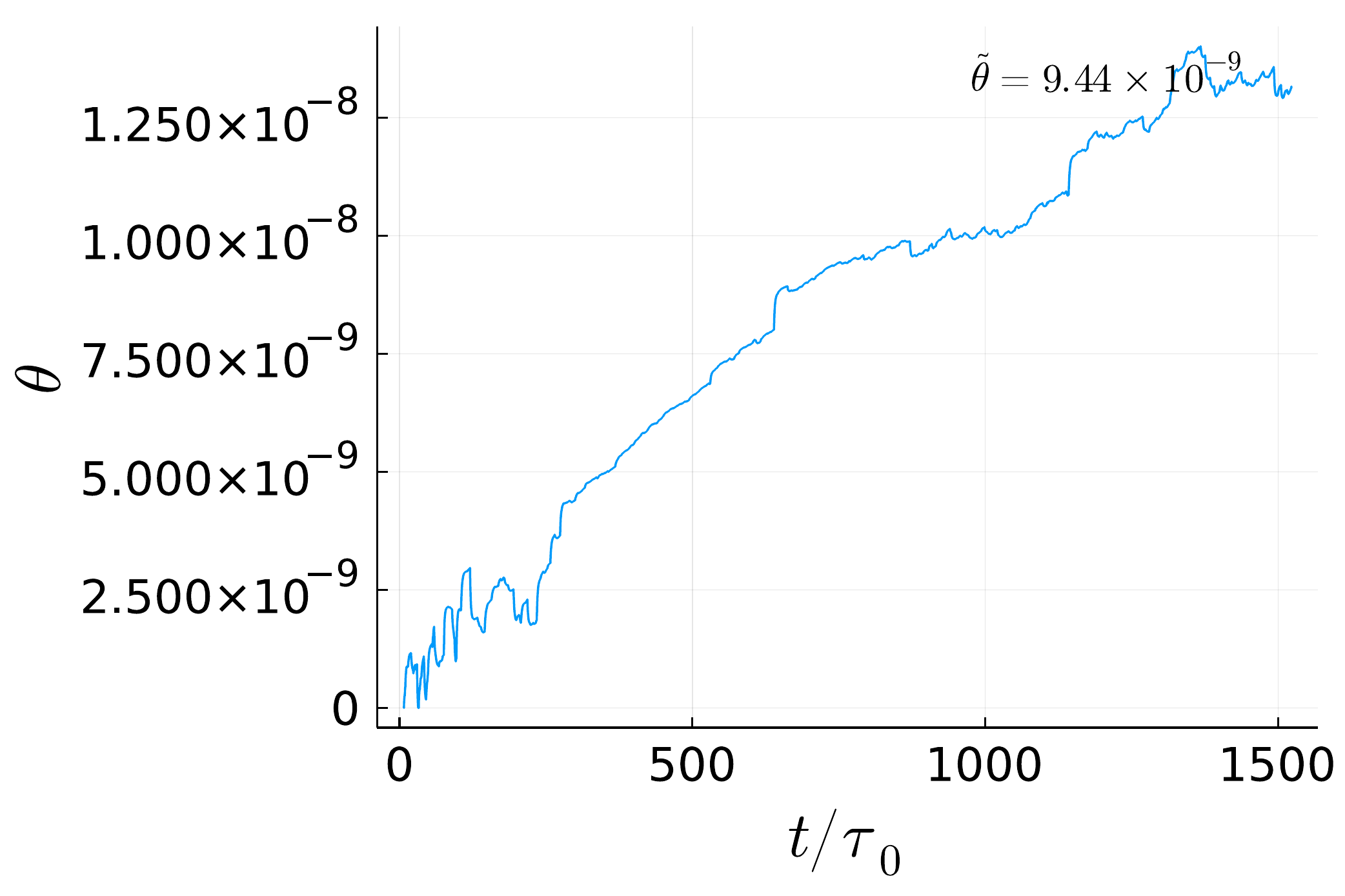}
\caption{Loss $L$ (left) and parameter $\theta$ (right) during optimization with learning rate $1\times10^{-10}$ when we update parameter~$\theta$ every 10 time steps. Median values are depicted as $\tilde{L}$ and $\tilde{\theta}$ respectively.}
\label{fig:GOY_ml_lr=1e-10_update_freq=100}
\end{figure}

We show the evolution of the loss~$L$ from Equation~\ref{eq:objective} and the learned parameter~$\theta$, when the optimization is run for longer time periods, namely~${\sim1500}$ large eddy turnover times. Figure~\ref{fig:GOY_ml_lr=1e-9_update_freq=1},~\ref{fig:GOY_ml_lr=1e-9_update_freq=10} and~\ref{fig:GOY_ml_lr=1e-9_update_freq=100} show results for learning rate $1\times10^{-9}$ when we update parameter~$\theta$ every 0.1, 1 or 10 time steps respectively. Figure~\ref{fig:GOY_ml_lr=1e-10_update_freq=1},~\ref{fig:GOY_ml_lr=1e-10_update_freq=10} and~\ref{fig:GOY_ml_lr=1e-10_update_freq=100} show results for a smaller learning rate of $1\times10^{-10}$ when we update parameter~$\theta$ every 0.1, 1 or 10 time steps respectively. We display the median values for the loss~$L$ and parameter~$\theta$ in the graphs as~$\tilde{L}$ and~$\tilde{\theta}$ respectively. The median value for $\theta$ lands relative consistently between $1.5\times 10^{-8}$ and $2.0\times 10^{-8}$, except for the last experiment (Figure~\ref{fig:GOY_ml_lr=1e-10_update_freq=100}), where it does not seem to have quite converged. As evident from Figures~\ref{fig:GOY_ml_lr=1e-10_update_freq=10} and~\ref{fig:GOY_ml_lr=1e-10_update_freq=100} a smaller learning rate ($1\times 10^{-10}$) in combination with larger intervals between updates can lead to a significantly slower discovery of the correct value for $\theta \approx 10^{-8}$.


\end{document}